\documentclass[smallextended, natbib]{svjour3}

\smartqed

\usepackage{mathptmx}

\usepackage{graphicx}
\usepackage{amsmath}
\usepackage{booktabs}
\usepackage{xspace}
\usepackage{tabularx}
\usepackage{url}
\usepackage{etoolbox}
\usepackage{booktabs,longtable}
\usepackage{hyphenat}

\newcommand{\CC}{{C\nolinebreak[4]\hspace{-.05em}\raisebox{.4ex}{\tiny\bf ++}}\xspace}
\newcommand{\pkg}[1]{#1} 
\newcommand{\proglang}[1]{#1} 
\newcommand{\code}[1]{#1} 

\newcommand*{\NONEWPAGE}{} 

\newtoggle{blind}
\togglefalse{blind} 

\newtoggle{structured}
\toggletrue{structured} 

\begin{document}

\title{Statistical Software for Psychology:
Comparing Development Practices Between CRAN and Other Communities} 

\author{
\iftoggle{blind}
{~}
{Spencer Smith \and Yue Sun \and Jacques Carette}
}

\institute{
\iftoggle{blind}
{~}
{
Spencer Smith \and Yue Sun \and Jacques Carette \at
Computing and Software Department\\
McMaster University\\
Hamilton, Ontario, Canada, L8S 4K1\\
\email{\{smiths,carette\}@mcmaster.ca}
}
}

\titlerunning{Stats for Psych, comparing development practices}
\authorrunning{Smith et al.}


\maketitle

\iftoggle{structured}
{
\abstract{{\bf Context}:  Different communities rely heavily on
  software, but use quite different software development practices.
{\bf Objective}:  We wanted to measure the state of the practice
in the area of statistical software for psychology to understand
how it compares to best practices.
{\bf Method}: We compared and ranked 30 software tools with respect 
to adherence to best software engineering practices on items that could
be measured by end-users.
{\bf Results}  We found that R packages use quite good practices,
that while commercial packages were quite usable, many aspects of their
development is too opaque to be measures, and that research projects
vary a lot in their practices.
{\bf Conclusion}  We recommend that more organizations adopt practices
similar to those used by CRAN to facilitate success, even for small teams.
We also recommend close coupling of source code and documentation, to 
improve verifiability.
}
}
{\abstract{This paper looks at the application of software engineering ideas to
  the development of software used in psychology.  Specifically, we compared the
  state of practice for developing statistical software for psychology by reviewing and
  ranking 30 software tools with respect to their adherence to software
  engineering best practices.  Our goal was to identify the most successful
  development approaches and areas for improvement.  We compared packages hosted
  by the Comprehensive \proglang{R} Archive Network (CRAN) to packages developed
  in other environments.  Our analysis showed that \proglang{R} packages use
  good software engineering practices to improve maintainability, reusability,
  understandability and visibility; commercial software packages tend to provide
  the best usability, but do not do as well on verifiability; and, software
  produced by research projects are not uniform in terms of their organization
  or consistency.  The software development process, tools and policies used by
  CRAN provide an infrastructure that facilitates success, even for small
  development teams.  On the other hand, \proglang{R} extensions, as well as
  code written in other languages, could improve their verifiability and
  understandability if tools were used to
  couple the source code and its documentation.}
}

\section[Introduction]{Introduction} \label{Introduction}
Best practices can be slow to propagate between disciplines.  This
paper attempts to address this problem between the fields of psychology and
software engineering.  In particular, we look at the state of practice for the
development of statistical software meant to be used in psychology%
\footnote{For brevity, we will abbreviate this to SSP for the rest of this paper.}.  
Developers of SSP, as
in other scientific domains, frequently develop their own software because
domain-specific knowledge is critical for the success of their applications
\citep{wilson-best-practices}.  However, these scientists are often self-taught
programmers and thus potentially unaware of software development best practices.  To
help remedy this situation, \citet{wilson-best-practices} provide general advice
for scientific software development.  We look at how well this advice is applied
in the specific scientific domain of SSP.  Our goal is to first understand what the
state of the practice is in SSP, and then provide advice as to what software
engineering practices would likely provide the biggest gains in perceived 
software quality, \emph{as measured by end-user perception}.

A first look at the state of practice for software in a specific scientific
community is provided by \citep{gewaltig-neuroscience}, for the domain of
computational neuroscience.  (A newer version of their paper is available
\citep{newneuro}, but we reference the original version, since its simpler
software classification system better matches our needs.)
\citet{gewaltig-neuroscience} provide a high level comparison of existing
neuroscience software, but little data is given on the specific metrics for
their comparison.  We build on the idea of studying software created and used by
a specific scientific community, while also incorporating detailed measures of
software qualities.  In this paper we use the term \emph{software qualities} as
used by software engineers to refer to properties of software such as
installability, reliability, maintainability, portability etc.  When we speak of
\emph{software qualities}, we mean the union of these properties.

Here we target 30 SSP packages, developed by different
communities using different models.  The packages were selected from a
combination of three external lists: \citet{pswiki:Online}, the National Council
on Measurement in Education \citet{psdb:Online}, and the Comprehensive
\proglang{R} Archive Network \citet{pscran:Online}.

Combining our own ideas with suggestions from \citet{wilson-best-practices} and
\citet{gewaltig-neuroscience}, we created a grading sheet to systematically
measure each package's qualities.  We used the Analytic Hierarchy Process (AHP)
\citep{AHP} to quantify the ranking between packages via pair-wise comparisons.
Our grading assumes that the software is intended to be user ready, as defined
by \citet{gewaltig-neuroscience}.  That is, the grading assumes that the
intention is for new users to be able to undertake their work without requiring
communication with the original developers.  In many cases a low ``grade''
should not be attributed to a deficiency in the software, but rather to the fact
that the overall goal was not user readiness, but rather research readiness
\citep{gewaltig-neuroscience}.  The overall quality target should be taken into
account when interpreting the final rankings.

Unlike \citet{gewaltig-neuroscience}, the authors of this study are not domain
experts.  Our aim is to analyze SSP with respect to software
engineering aspects only. Due to our lack of domain knowledge, algorithms and
background theory will not be discussed, and packages will not be judged
according to the different functionalities they provide.  We reach our
conclusions through a systematic and objective grading process, which
incorporates some experimentation with each package.

Others have looked at issues surrounding the engineering of scientific software.
Of particular relevance is \citet{Heaton2015207}, which looks at $12$ different
software engineering practices across $43$ papers that examine software
development as performed by scientists.  The software engineering practices are
grouped as \emph{development workflow}, consisting of design issues, lifecycle
model, documentation, refactoring, requirements, testing, and verification and
validation; and \emph{infrastructure}, consisting of issue tracking, reuse,
third-party issues and version control.  These, naturally, have significant
overlap with software qualities, as these practices are \emph{supposed} to
improve these qualities.  Even though this is a survey of practices, and one
would expect that this would be biased towards success stories and thus fairly
good practices, what emerges is different.  In other words, even among success
stories, the state of the practice is rather mixed.  This further motivates us
to look at the state of the practice of SSP projects ``from the outside'', and
thus picking from a (hopefully) wider cross-section of projects.  Another
relevant study is \citet{Kanewala20141219}, where the authors systematically
reviewed $62$ studies of relevance to software testing, motivated by the
increasing number of paper retractions traceable to software faults.  The main
conclusion is that the cultural difference between scientist developers and
software engineers, coupled with issues specific to scientific software, makes
testing very difficult.  We are gratified that this independent study justifies
both our choice to not do our own independent testing, as well as the idea that
investigating the ``software engineering maturity level'' of particular domains
is likely to find non-trivial variations.

Background information is provided in the first section below.  This is followed
by the experimental results and basic comparisons between packages, along with
information on how the software developed by the CRAN community compares to
SSP developed by other communities.

\section[Background]{Background} \label{Sec_Background}

This section covers the process used for our study and the rationale behind it.
We also introduce the terms and definitions used to construct the software
quality grading sheet and the AHP technique, which we used to make the
comparisons.  The process was,

\begin{enumerate}

\item Choose a domain where scientific computing is important.  Here we chose
  SSP because of its active software community, as evidenced by the
  list of open source software summarized by \citet{psdb:Online}, and a large
  number of \proglang{R} packages hosted by \citet{pscran:Online}.

\item Pick 30 packages from authoritative lists.  For SSP, this is a
  combination of NCME, CRAN and Wikipedia lists mentioned previously.  We picked
  15 \proglang{R} packages hosted by CRAN, and another 15 developed by other
  communities.  Packages common to at least two of the lists (15 out 30) were
  selected first, with the remaining selected randomly.

\item Build our grading sheet.  The template used is given in
  Appendix~\ref{app:full-template} and is available at:
  \url{https://github.com/adamlazz/DomainX}.

\item Grade each software.  Our main goal here is to remain objective.  To
  ensure that our process is reproducible, we asked different people to grade
  the same package.  The result showed that different people's standards for
  grading vary, but the overall ranking of software remained the same, since the
  overall ranking is based on relative comparisons, and not absolute grades.

\item Apply AHP on the grading sheet to reduce the impact of absolute grade
  differences.

\item Analyze the AHP results, using a series of different weightings, so that
  conclusions and recommendations can be made.

\end{enumerate}

\subsection[Types of software]{Categories and Status} \label{typeofsoftware}

The development models of each package fell into three categories:

\begin{enumerate}

\item Open source: ``Computer software with its source code made available and
  licensed so the copyright holder provides the rights to study, change and
  distribute the software to anyone for any purpose'' \citep{license}.

\item Freeware: ``Software that is available for use at no monetary
  cost, but with one or more restricted usage rights such as source code being
  withheld or redistribution prohibited'' \citep{freeware}.

\item Commercial: ``Computer software that is produced for sale or that
  serves commercial purposes'' \citep{Dictionary.com2014}.

\end{enumerate}

\noindent The \emph{status} of each project is said to be \textbf{Alive} if the
package, related documentation, or web site has been updated within the last 18
month; \textbf{Dead} if the last update was $18$ months ago or longer;
\textbf{Unclear} if last release information could not be easily derived
(denoted ? in tables).

\subsection[Software qualities]{Software qualities} \label{sec:qual}

We use the software engineering terminology from \citet{ghezzi-se} and best
practices from \citet{wilson-best-practices} to derive our terms and measures.
We measure items of concern to both end users and developers.  Since the
terminology is not entirely fixed across the software engineering literature, we
provide the definitions we will use throughout.  The qualities are presented in
(roughly) the order we measured them.  Where relevant, information on how each
quality was measured is given.

\begin{itemize}

\item \textit{Installability} is a measure of the ease of software installation.
  This is largely determined by the quantity and quality of installation
  information provided by developers.  Good installability means detailed and
  well organized installation instructions, with less work to be done by users
  and automation whenever possible.

\item \textit{Correctness and Verifiability} are related to how much a user can
  trust the software.  Software that makes use of trustworthy libraries (those
  that have been used by other packages and tested through time) can bring more
  confidence to users and developers than self-developed
  libraries \citep{Dubois2005}.  Carefully documented specifications should also
  be provided.  Specifications allow users to understand the background theory
  for the software, and its required functionality.  Well explained examples
  (with input, expected output and instructions) are helpful too, so that users
  can verify for themselves that the software produces the same result as
  expected by the developers.

\item \textit{Reliability} is based on the dependability of the software.
  Reliable software has a high probability of meeting its stated requirements
  under a given usage profile over a given span of time.

\item \textit{Robustness} is defined as whether a package can handle unexpected
  input.  A robust package should recover well when faced with unexpected input.

\item \textit{Performance} is a measure of how quickly a solution can be found
  and the resources required for its computation.  Given the constraint that we
  are not domain experts, in the current context we are simply looking to see
  whether there is evidence that performance is considered.  Potential evidence
  includes signs of use of a profiler, or other performance data.

\item \textit{Usability} is a measure of how easy the software is to use.  This
  quality is related to the quality and accessibility of information provided by
  the software and its developers.  Good documentation helps with usability.
  Some important documents include a user manual, a getting started tutorial,
  and standard examples (with input, expected output and instructions).  The
  documentation should facilitate a user quickly familiarizing themselves with
  the software.  The GUI should have a consistent look and feel for its
  platform.  Good visibility \citep{norman} can allow the user to find the
  functionality they are looking for more easily.  A good user support model
  (e.g.\ forum) is beneficial as well.

\item \textit{Maintainability} is a measure of the ease of correcting and
  updating the software.  The benefits of maintainability are felt by future
  contributors (developers), as opposed to end users.  Keeping track of version
  history and change logs facilitates developers planning for the future and
  diagnosing future problems.  A developer's guide is necessary, since it
  facilitates new developers doing their job in an organized and consistent
  manner.  Use of issue tracking tools and concurrent version system is a good
  practice for developing and maintaining software
  \citep{wilson-best-practices}.

\item \textit{Reusability} is a measure of the ease with which software code can
  be used by other packages.  In our project, we consider a software package to
  have good reusability when part of the software is used by another package and
  when the API (Application Program Interface) is documented.

\item \textit{Portability} is the ability of software to run on different
  platforms.  We examine a package's portability through developers' statement
  from their web site or documents, and the success of running the software on
  different platforms.

\item \textit{Understandability (of the code)} measures the quality of
  information provided to help future developers with understanding the
  behavior of the source code.  We surface check the understandability by
  looking at whether the code uses consistent indentation and formatting style,
  if constants are not hard coded, if the code is modularized, etc.  Providing a
  code standard, or design document, helps people become familiar with the code.
  The quality of the algorithms used in the code is not considered here.

\item \textit{Interoperability} is defined as whether a package is designed to
  work with other software, or external systems.  We checked whether that kind
  of software or system exists and if an external API document is provided.

\item \textit{Visibility/Transparency} is a measure of the ease of examining the
  status of the development of a package.  We checked whether the development
  process is defined in any document.  We also record the examiner's overall
  feeling about the ease of accessing information about the package.  Good
  visibility allows new developers to quickly make contributions to the project.

\item \textit{Reproducibility} is a measure of whether related information, or
  instructions, are given to help verify a products' results
  \citep{davison-reproducibility}.  Documentation of the process of verification
  and validation is required, including details of the development and testing
  environment, operating system and version number.  If possible, test data and
  automated tools for capturing the experimental context should be provided.

\end{itemize}

Our grading sheet, as shown in Appendix~\ref{app:full-template}, is derived from
these qualities.  Installability, for example was determined by asking the
questions shown in Table~\ref{table:example}, where \textit{Unavail} means that
uninstallation is not available, and the superscript $^*$ means that this
response should be accompanied by explanatory text.

\begin{table}
  \caption{Measuring installability}
  \label{table:example} 
  \begin{tabular}{l}
    \toprule
    Question (Allowed responses)\\
    \midrule
    Are there installation instructions? (Yes/ No)\\
    Are the installation instructions linear? (Yes/ No)\\
    Is there something in place to automate the installation? (Yes$^*$/ No)\\
    Is there a specified way to validate the installation, such as a test suite? (Yes$^*$/ No)\\
    How many steps were involved in the installation? (Number)\\
    How many packages need to be installed before or during installation?
    (Number)\\
    Run uninstall, if available. Were any obvious problems caused? (Unavail/Yes$^*$/ No)\\
    \bottomrule
  \end{tabular}
\end{table}

\subsection[Analytic Hierarchy Process]{Analytic Hierarchy Process
  (AHP)} \label{AHP}

``The AHP is a decision support tool which can be used to solve complex decision
problems.  It uses a multi-level hierarchical structure of objectives, criteria,
subcriteria, and alternatives.  The pertinent data are derived by using a set of
pairwise comparisons.  These comparisons are used to obtain the weights of
importance of the decision criteria, and the relative performance measures of
the alternatives in terms of each individual decision criterion''
\citep{trianta-ahp}.  By using AHP, we can compare between qualities and
packages without worrying about different scales, or units of
measurement.

Generally, by using AHP, people can evaluate $n$ options with respect to $m$
criteria.  The criteria can be prioritized, depending on the weight given to
them.  An $m \times m$ decision matrix is formed where each entry is the
pair-wise weightings between criteria.  Then, for each criterion, a pairwise
analysis is performed on each of the options, in the form of an $n\times n$
matrix $a$. (There is a matrix $a$ for each criteria).  It is formed through
using a pair-wise score between options as each entry.  The entry of the upper
triangle of $a$ is scaled between one to nine, defined as in \citet{Saaty1990}.


Here we compared 30 packages ($n = 30$) with respect to the 13 qualities
($m = 13$) mentioned previously.  Overall quality judgements will depend on the
context in which each package is meant to be used.  To approximate this, we
experimented with different weights for each property.

We capture a subjective score, from one to ten, for each package for each
criteria through our grading process.  To turn these into pair-wise scores, one
starts with two scores $A$ and $B$ (one for each package), and the result for
$A$ versus $B$ is:
\[
\begin{cases}
\max\{9, A - B + 1\} & A \geq B \\
1 / \min\{1, B - A + 1\} & A < B
\end{cases}
\]

For example, if installability is measured as an $8$ for package $A$ and a $2$
for package $B$, then the entry in $a$ corresponding to $A$ versus $B$ is
$7$, while that of $B$ versus $A$ is $1/7$.  The implication
is that installing $A$ is much simpler than installing $B$.

\section[Experimental results and discussion]{Experimental
  Results} \label{result}

We briefly introduce the $30$ packages.  Next we present the AHP results by
discussing trends for each quality and then looking at the final rankings,
assuming both equal and non-equal weights.  The detailed results are in
Appendix~\ref{app:SummaryMeasurements} and available on-line at
\url{https://github.com/adamlazz/DomainX}.

\subsection[Selection of Software]{Packages}

Summary information on the $30$ packages is presented in Tables
\ref{table:r}--\ref{table:commercial}.  In particular,

\begin{itemize}
\item 19 packages are open source; 8 are freeware; and 3 are commercial.
\item 15 are developed using \proglang{R} (Table~\ref{table:r}) and maintained
  by the CRAN community; 12 are from university projects, or developed by
  research groups (Table~\ref{table:research}); and 3 are developed by companies
  for commercial use (Table~\ref{table:commercial}).
\item 3 projects use \proglang{\CC}; 2 use \proglang{Java}; 2 use
  \proglang{Fortran}, and 1 uses \proglang{BASIC}.  The programming language of
  the remaining 7 is not mentioned by the developers.
\item All the packages from CRAN are alive, using the definition given in the
  background section.  Two of the commercial software product are alive and one
  is unclear.  For the rest of software packages (mostly from university
  projects or research groups), 6 are alive, 5 are dead and the last is unclear.
\end{itemize}

\begin{table}[ht]
  \caption{CRAN (\proglang{R}) packages}
  \label{table:r}  
  \begin{tabular}{lccccc} 
    \toprule 
    Name & Released & Updated & Status & Source & Lang.\\[0.5ex] 
    \midrule 
    \pkg{eRm} \citep{erm} & 2007 & 2014 & Alive & Available & \proglang{R}\\

    \pkg{Psych} \citep{Psych} & 2007 & 2014& Alive& Available& \proglang{R}\\ 
    \pkg{mixRasch} \citep{mixRasch} & 2009 & 2014& Alive& Available& \proglang{R}\\ 
    \pkg{irr} \citep{irr} & 2005 & 2014& Alive& Available& \proglang{R}\\  
    \pkg{nFactors} \citep{nFactors} & 2006 & 2014& Alive& Available& \proglang{R}\\ 
    \pkg{coda} \citep{coda} & 1999 & 2014& Alive& Available& \proglang{R}\\ 
    \pkg{VGAM} \citep{vgam} & 2006 & 2013& Alive& Available& \proglang{R}\\ 
    \pkg{TAM} \citep{tam} & 2013 & 2014& Alive& Available& \proglang{R}\\  
    \pkg{psychometric} \citep{psychometricp} & 2006 & 2013& Alive& Available& \proglang{R}\\ 
    \pkg{ltm} \citep{ltm} & 2005 & 2014& Alive& Available& \proglang{R}\\  
    \pkg{anacor} \citep{anacor} & 2007 & 2014& Alive& Available& \proglang{R}\\ 
    \pkg{FAiR} \citep{FAiR} & 2008 & 2014& Alive& Available& \proglang{R}\\ 
    \pkg{lavaan} \citep{lavaan} & 2011 & 2014& Alive& Available& \proglang{R}\\  
    \pkg{lme4} \citep{lme4} & 2003 & 2014& Alive& Available& \proglang{R}\\ 
    \pkg{mokken} \citep{mokken} & 2007 & 2013& Alive& Available& \proglang{R}\\ 
    [1ex] 
    \bottomrule
  \end{tabular}  
\end{table}

\begin{table}[h]  
  \caption{Other research group projects}
  \label{table:research}
  \begin{tabular}{l c c c c c} 
    \toprule
    Name & Released & Updated & Status & Source & Lang.\\[0.5ex] 
    \midrule 
    \pkg{ETIRM} \citep{ETIRM} & 2000 & 2008 & Dead & Available & \proglang{\CC}\\ 
    \pkg{SCPPNT} \citep{SCPPNT} & 2001 & 2007& Dead& Available& \proglang{\CC}\\
    \pkg{jMetrik} \citep{jMetrik} & 1999 & 2014 & Alive & Not&
                                                              \proglang{Java}\\
    \pkg{ConstructMap} \citep{ConstructMap} & 2005 & 2012& Dead& Not&
                                                                     \proglang{Java}\\ 
    \pkg{TAP} \citep{TAP} & ? & ?& ?& Not& ?\\ 
    \pkg{DIF-Pack} \citep{DIF} & ? & 2012& Alive& Available&
                                                            \proglang{Fortran}\\ 
    \pkg{DIM-Pack} \citep{DIM} & ? & 2012& Alive& Available&
                                                            \proglang{Fortran}\\ 
    \pkg{ResidPlots-2} \citep{ResidPlots-2} & ? & 2008& Dead& Not&
                                                                  ?\\ 
    \pkg{WinGen3} \citep{WinGen3} & ? & 2013& Alive& Not& ?\\ 
    \pkg{IRTEQ} \citep{IRTEQ} & ? & 2011& Dead& Not& ?\\ 
    \pkg{PARAM} \citep{PARAM} & ? & 2012& Alive& Not&
                                                     \proglang{BASIC}\\ 
    \pkg{IATA} \citep{IATA} & ? & 2014& Alive& Not & ?\\ 
    [1ex]
    \bottomrule
  \end{tabular} 
\end{table}

\begin{table}[h]
  \caption{Commercial packages}
  \label{table:commercial} 
  \begin{tabular}{l c c c c c c} 
    \toprule 
    Name & Released & Updated & Status & Source & Lang.\\[0.5ex] 
    \midrule
    \pkg{MINISTEP} \citep{ministep} & 1977 & 2014 & Alive & Not & ?\\
    \pkg{MINIFAC} \citep{minifac} & 1987 & 2014& Alive& Not& ?\\ 
    \pkg{flexMIRT} \citep{flexMIRT} & ? & ? & ? & Not& \proglang{\CC}\\
    [1ex]
    \bottomrule
  \end{tabular} 
\end{table}

\subsection[AHP results]{AHP results}

We explain the AHP results for each quality.  In the charts, blue bars (slashes)
are for packages hosted by CRAN\footnote{As all relevant \proglang{R} packages
  are hosted by CRAN, we will drop this qualifier for the rest of this
  paper.}, green bars (horizontal lines) for research groups projects and red
bars (vertical lines) for commercial software.

\subsubsection{Installability} \label{Installability} We
installed each package on a clean virtual machine.  We did this to ensure we
used a clean environment for each installation, to not create bias for software
tested later.  We also checked for installation instructions, and whether these
instructions are organized linearly.  Automated tools for installation and test
cases for validation are preferred.  The number of steps during installation and
the number of external libraries required was counted.  At the end of the
installation and testing, if an uninstaller is available, we ran it to see if it
completed successfully.  The AHP results are in Figure~\ref{fig:installability}.
Some key findings are as follows:

\begin{figure}[ht]
\centering
\includegraphics[width=0.9\textwidth]{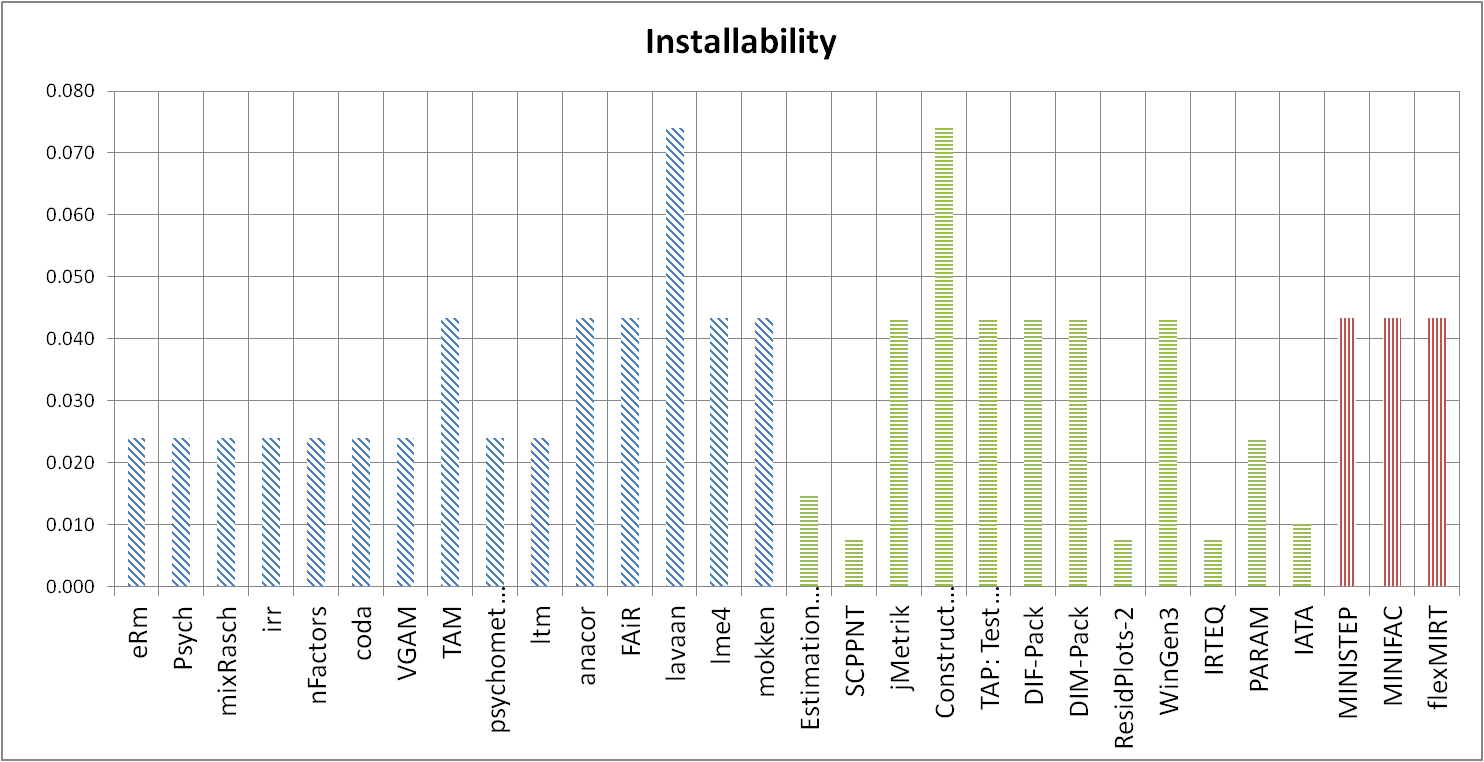}
\caption{AHP result for installability}
\label{fig:installability}
\end{figure}

\begin{itemize}

\item \proglang{R} packages: The CRAN community provides general installation
  instructions \citep{craninstall} for all packages it maintains. However, in
  some cases links to the general instructions are not given in the package page,
  which may cause confusion for beginner users.  The following packages
  addressed this problem by providing detailed installation information on their
  own web site: \pkg{TAM}, \pkg{anacor}, \pkg{FAiR}, \pkg{lavaan}, \pkg{lme4}
  and \pkg{mokken}. All the packages installed easily and automatically. Most of
  the packages have dependencies on other \proglang{R} packages, but the
  required packages can be found and installed automatically.  The
  uninstallation process is as easy as the installation process. A drawback of
  \proglang{R} packages for installability is that none of the packages provide
  a standard suite of test cases specifically for the verification of
  installation.  \pkg{lavaan} provide a simple test example, without output
  data, which does show some consideration toward verification of the
  installation process.

\item Research group projects: The results of installability in the research
  group projects are uneven.  Several are similar to \proglang{R} packages, but
  some rank lower because no installation instructions are given (5 out of 12),
  or the given instruction is not linearly organized (one). The developers may
  think that the installation process is simple enough that there is no need for
  installation instructions. However, even for a simple installation, it is good
  practice to have documentation to prevent trouble for new users. Another
  common problem is the lack of a standard suite of test cases, only
  \pkg{ConstructMap} and \pkg{PARAM} showed consideration of this issue.

\item Commercial software: They have well organized installation instructions,
  automated installers and only a few tasks needed to be done manually. However,
  commercial software tends to have the same problem as the \proglang{R}
  packages -- in no instance is a standard test suite provided specifically for
  the verification of the installation.

\end{itemize}

\subsubsection{Correctness and Verifiability}

\begin{figure}[ht]
\centering
\includegraphics[width=0.9\textwidth]{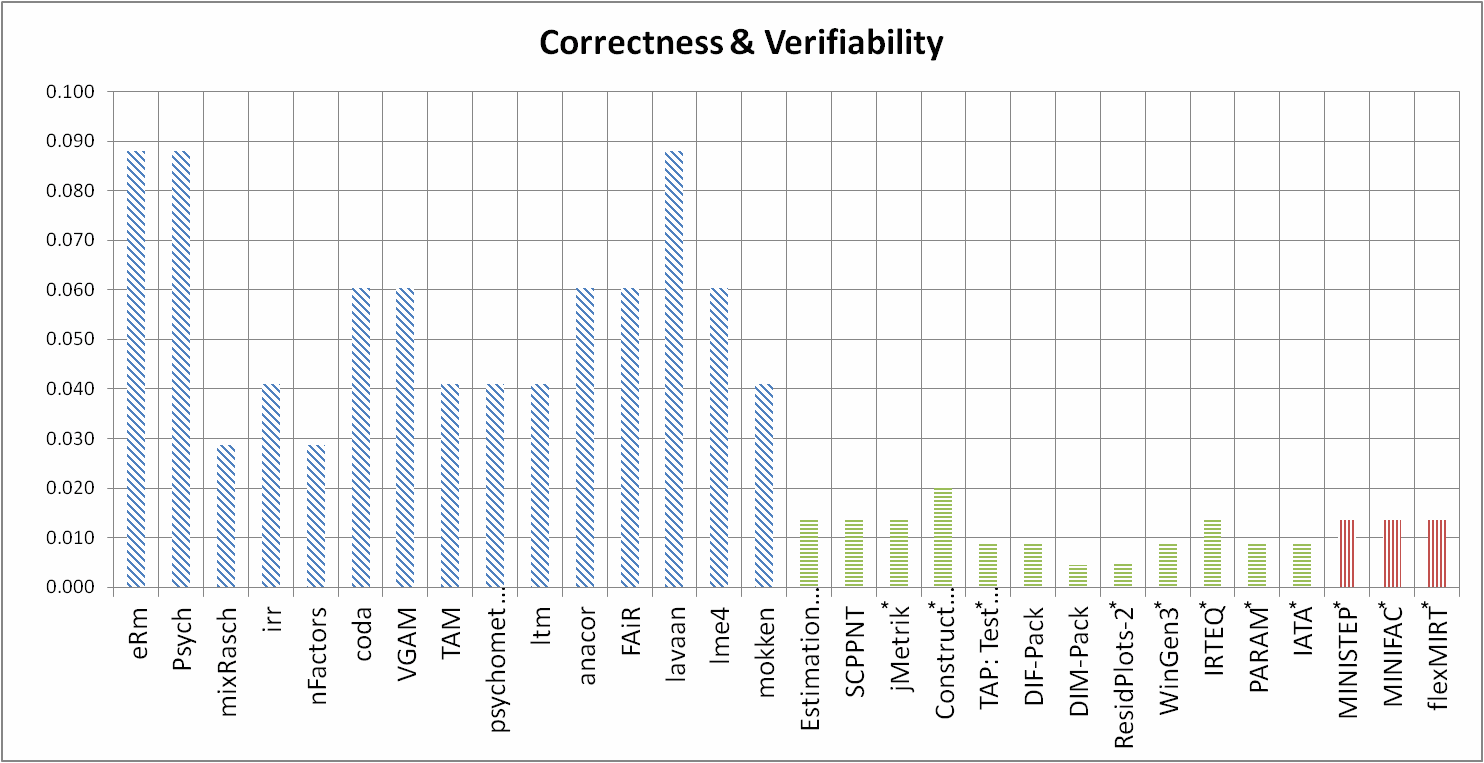}
\caption{AHP result for correctness and verifiability (close source indicated by *)}
\label{fig:correctness} 
\end{figure}

We checked for evidence of trustworthy libraries and a requirements
specification.  With respect to the requirements specification, we did not
require the document to be written in a strict software engineering style -- it
was considered adequate if it described the relevant mathematics.  We tried the
given examples, if any, to see if the results matched the expected output.  The
results are shown in Figure~\ref{fig:correctness}.  In particular, we observe
that

\begin{itemize}

\item \proglang{R} packages: For $10$ of the $15$, there are either technical
  reports or published papers about the mathematics used by the package.  For
  instance, \pkg{lme4}, \pkg{eRm} and \pkg{mokken} are covered in a special
  volume (number 20) on SSP from the Journal of
  Statistical Software (JSS) \citep{deLeeuwAndMair2007}.  A later special volume
  (number 48) of JSS covers further \proglang{R} extensions,
  including \pkg{lavaan} \citep{Rosseel2012}.  The software has consistent
  documentation because \proglang{R} extensions must satisfy the CRAN Repository
  policy, which includes standardized interface documentation through
  \proglang{Rd} (\proglang{R} documentation) files.  \proglang{Rd} is a markup
  language that can be processed to create documentation in \proglang{\LaTeX},
  \proglang{HTML} and text formats \citep{RCoreTeam2014}.  Although not required
  by the CRAN policy, some of the \proglang{R} extensions also include
  vignettes, which provide additional information in a more free format than the
  \proglang{Rd} documentation allows.  Vignettes can act as user manuals,
  tutorials, and extended examples.  Many vignettes are written with
  \proglang{Sweave} \citep{Leisch2002}, which is a Literate Programming (LP)
  \citep{Knuth1984} tool for \proglang{R}.  All the \proglang{R} packages have
  examples about how to use the software, but $9$ did not provide expected
  output data; we encountered a small precision difference when we ran \pkg{ltm}
  and compared with the expected results.  Many packages rely on other packages
  (which can be seen as CRAN provides package dependencies as well as reverse
  dependencies); such reuse not only eases development burden, but reused
  packaged tend to be generally more trustworthy than newly developed libraries.

\item Research group projects: None provide requirements specifications nor
  reference manuals.  Examples are given in most case ($11$ of $12$), but $9$ of
  those did not provide input or output data.  Only two mentioned that standard
  libraries were used.

\item Commercial: Much less information (especially as compared to \proglang{R})
  was provided for building confidence in correctness and verifiability.
  Neither did they provide requirement specification documents, or reference
  manuals.  As this is commercial software, it may be the case that the
  companies believe that their proprietary algorithms given them added value; we
  should not prematurely conclude that these packages are not correct or
  verifiable.  On the plus side, all these packages provide standard examples
  with relevant output and all the calculated results produced from our machine
  matched the expected results.  None of the selected packages mentioned whether
  they used existing popular libraries.

\end{itemize}

Little evidence of verification via testing was found among the three classes of
software.  The notable exception to this is the diagnostic checks done by CRAN
on each of the \proglang{R} extensions.  To verify that the extensions will
work, the \proglang{R} package checker, \code{R CMD check}, is used.  The tests
done by \code{R CMD check} include \proglang{R} syntax checks, \proglang{Rd}
syntax and completeness checks and, if available, example checks, to verify that
the examples produce the expected output \citep{RCoreTeam2014}.  These checks
are valuable, but they focus more on syntax than on semantics.  The \proglang{R}
community has not seemed to fully embrace automated testing, since common
development practices of \proglang{R} programmers do not usually include
automatically re-running test cases \citep{Wickham2011}.

Of the three classes of software, \proglang{R} packages provides the most
complete and consistent documentation, but there also seems to be a missed
opportunity here.  If there is a drawback to the \proglang{R} documentation, it
is that the documents only assist with the use of the tools, not their
verification.  In other words, although LP is used for user documentation, it is
not used for the implementation itself, so that documented code can be more
easily verified.  \citet{Nedialkov2010} is a good example of the improved
verifiability one can achieve by maintaining the code and documentation together
in one source; the CRAN community does not appear to follow this approach.

\subsubsection{Reliability (Surface)}

We checked rudimentary reliability by attempting to run the package after
installing it.  In particular, if there was a tutorial for the package, we tried
to run through its examples to see if we obtained the same results.  This can
only be considered a surface measure of reliability, as we did not conduct any
domain-specific checks.

\begin{figure}[ht]
\centering
\includegraphics[width=0.9\textwidth]{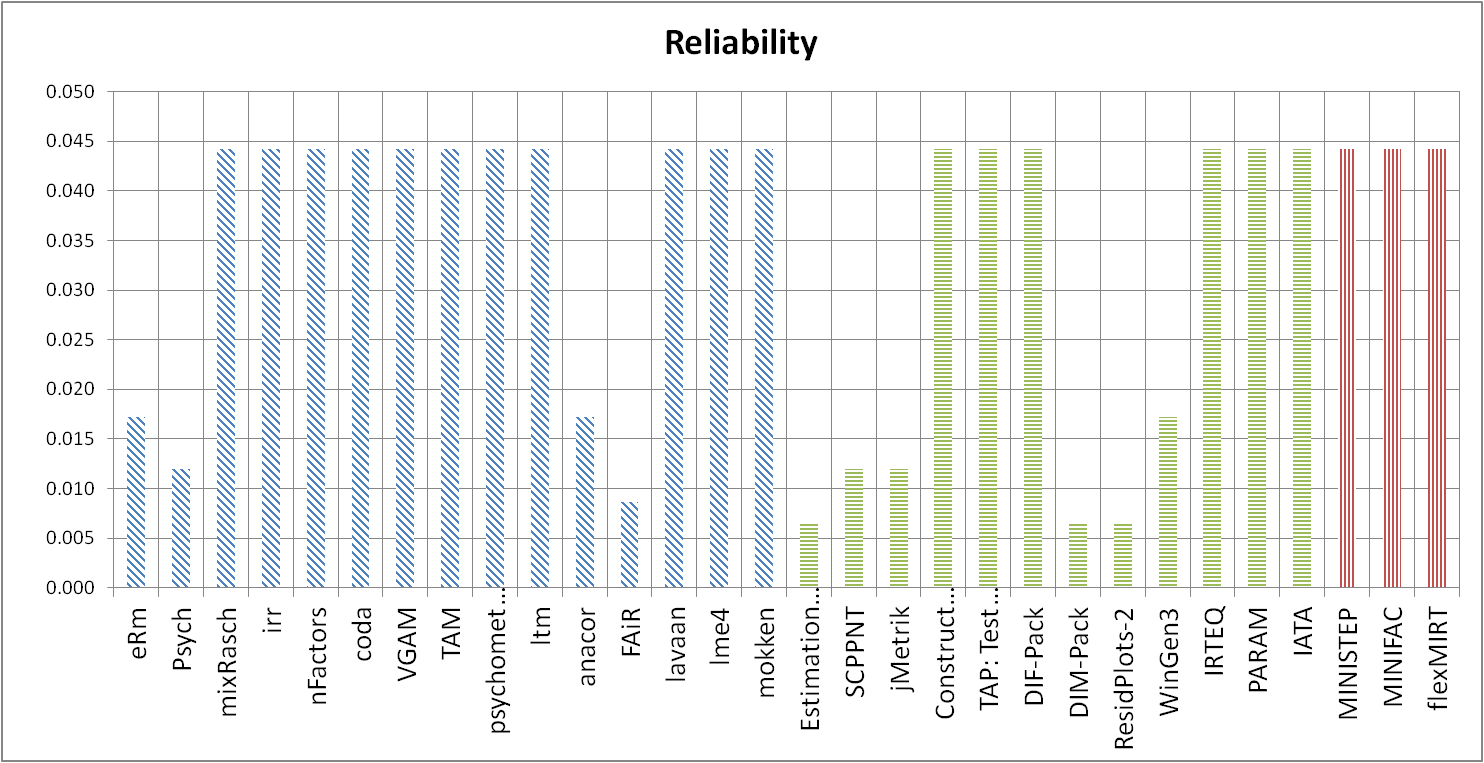}
\caption{AHP result for reliability}
\label{fig:reliability}
\end{figure}

\begin{itemize}

\item \proglang{R} packages: For $11$ packages, there were no problems during
  installation and initial tutorial testing. There were small installation
  problems for \pkg{eRm}, \pkg{Psych}, \pkg{anacor}, \pkg{FAiR} because the
  instructions were not up to date with the software.  The problems included
  dead URLs and a lack of required packages.  Our suggestion is that developers
  should maintain their install instructions, or point users to the general
  instruction given by CRAN.

\item Research group projects: For half of these, we found no problems.  The
  other half suffered from problems like Makefile errors (\pkg{SCPPNT}), old
  instructions (\pkg{WinGen3}), or an absence of instructions.

\item Commercial software: There were no problem during installation and initial
  tutorial testing when using the given instructions.

\end{itemize}

\subsubsection{Robustness (Surface)}

We checked robustness by providing the software with ``bad'' input.  We wanted
to know how well they deal with situations that the developers may not have
anticipated.  For instance, we checked whether they handle garbage input (a
reasonable response may be an appropriate error message) and whether they
gracefully handle text input files where the end of line character follows a
different convention than expected.  Like reliability, this is only a surface
check.  With reference to Figure~\ref{fig:robustness}, we have the following
remarks:

\begin{figure}[ht!]
\centering
\includegraphics[width=0.9\textwidth]{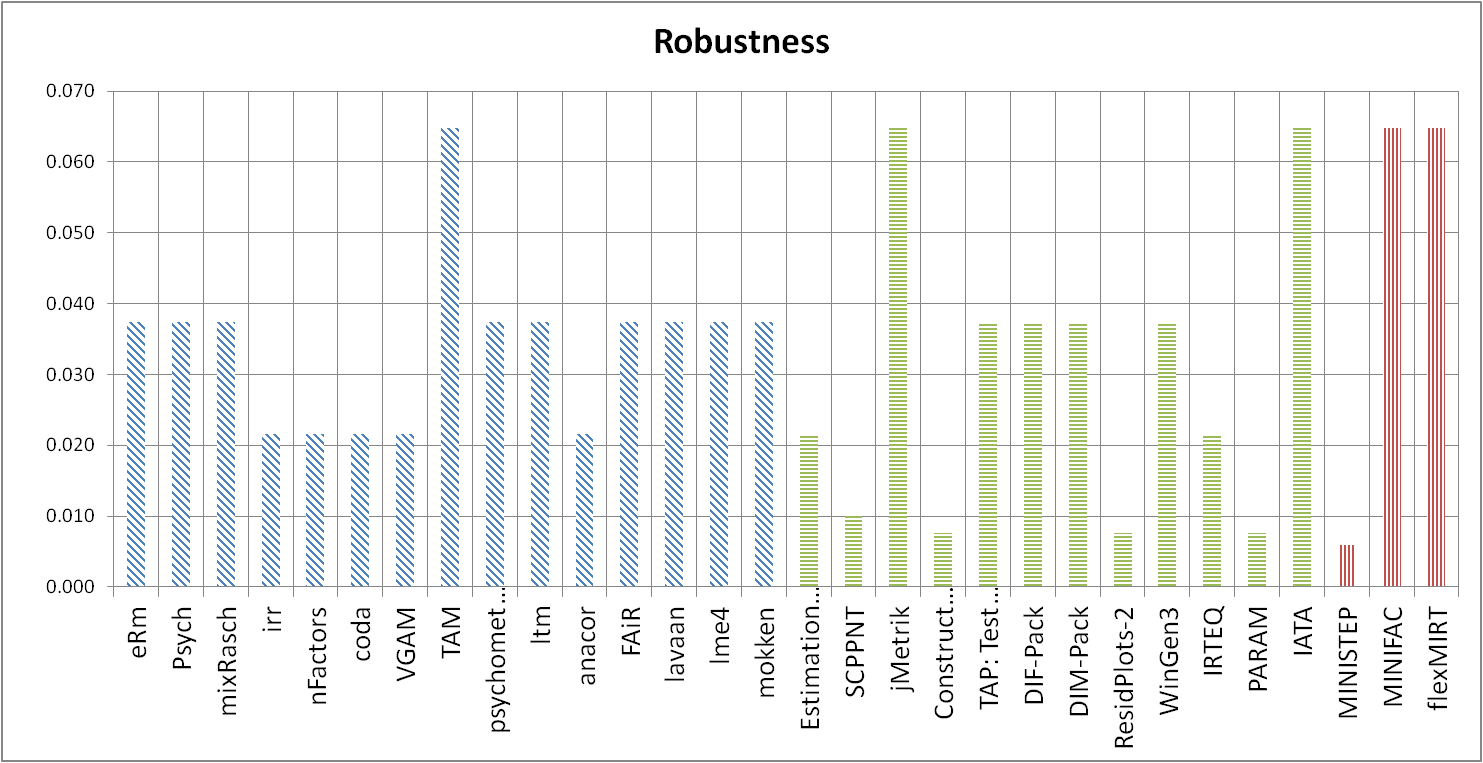}
\caption{AHP result for robustness}
\label{fig:robustness} 
\end{figure}

\begin{itemize}
\item \proglang{R} packages: $10$ handle unexpected input reasonably; they
  provide information or warnings when the user enters invalid
  data. \proglang{R} packages do not use text files as input files; therefore, a
  change of format in a text file is not an issue.
\item Research group projects: $7$ performed well.  The rest had problems like
  an inability to handle when the input file is not present, or when it does not
  have the designated name.
\item Commercial software: two did well, but \pkg{MINISTEP} did not issue a
  warning when using an invalid format for an input -- and the software crashed.
\end{itemize}

\subsubsection{Performance (Surface)}

No useful evidence or trends were obtained for performance; therefore, no
comparison can be made. 

\subsubsection{Usability (Surface)}

We checked usability mainly by looking at the documentation.  Better usability
means the users get more help from the developers.  We checked for a getting
started tutorial, for a fully worked example and for a user manual.  Also, we
looked for a reasonably well designed GUI, a clear declaration of the expected
user characteristic, and for information about the user support model.  The
results are shown in Figure~\ref{fig:usability}.  We observed the following:

\begin{figure}[ht!]
\centering
\includegraphics[width=0.9\textwidth]{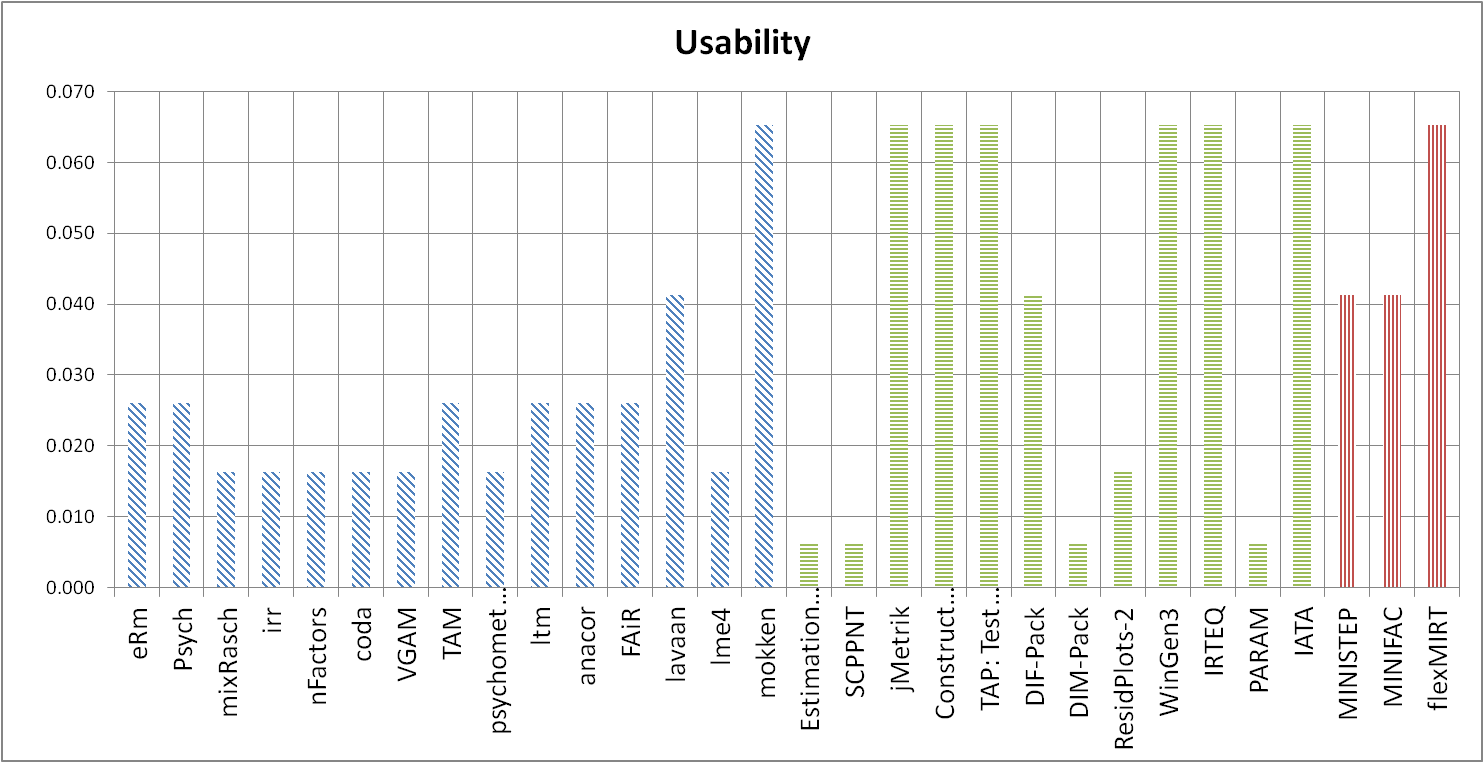}
\caption{AHP result for usability}
\label{fig:usability} 
\end{figure}

\begin{itemize}

\item \proglang{R} packages: Thanks to the CRAN repository policy, \proglang{Rd}
  files and \proglang{Sweave} vignettes, users are provided with complete and
  consistent documentation.  With respect to usability, two notably good
  examples are \pkg{mokken} and \pkg{lavaan}.  They give detailed explanations
  for their examples and provide well-organized user manuals. However,
  \proglang{R} packages have a common drawback -- only a few (4 out of 15)
  provide getting started tutorials.  The common user support model is an email
  address, with only two (\pkg{eRm} and \pkg{anacor}) providing a discussion
  forum as well.

\item Research group projects: There was great inconsistency here; a few
  packages (\pkg{IATA}, \pkg{ConstructMap}, \pkg{TAP: Test Analysis Program}
  come to mind) provided the best examples for others to follow.

\item Commercial software: Commercial software did better than the \proglang{R}
  packages here. They all have getting started tutorials, explanations for their
  examples and good user manuals.  The main drawback is that their GUIs do not
  have the usual look and feel for the platforms they are on.  For the user
  support model, \pkg{MINISTEP} and \pkg{MINIFAC} provide user forums and a
  feedback section, in addition to email support.

\end{itemize}

\subsubsection{Maintainability}

We looked for evidence that the software has actually been maintained and that
consideration was given to assisting developers with maintaining the software.
We looked for a history of multiple versions, documentation on how to contribute
or review code, and a change log.  In cases where there was a change log, we
looked for the presence of the most common types of maintenance (corrective,
adaptive or perfective).

We were also concerned about the tools that the developers used.  For instance,
what issue tracking tool was employed and does it show when major bugs were
fixed?  Since it is important for all scientific software
\citep{wilson-best-practices}, we also looked to see which versioning system is
in use.  Effort toward clear, non-repetitive code is considered as evidence for
maintainability. The results are shown in Figure~\ref{fig:maintainability}, and
we can highlight the following:

\begin{figure}[ht!]
\centering
\includegraphics[width=0.9\textwidth]{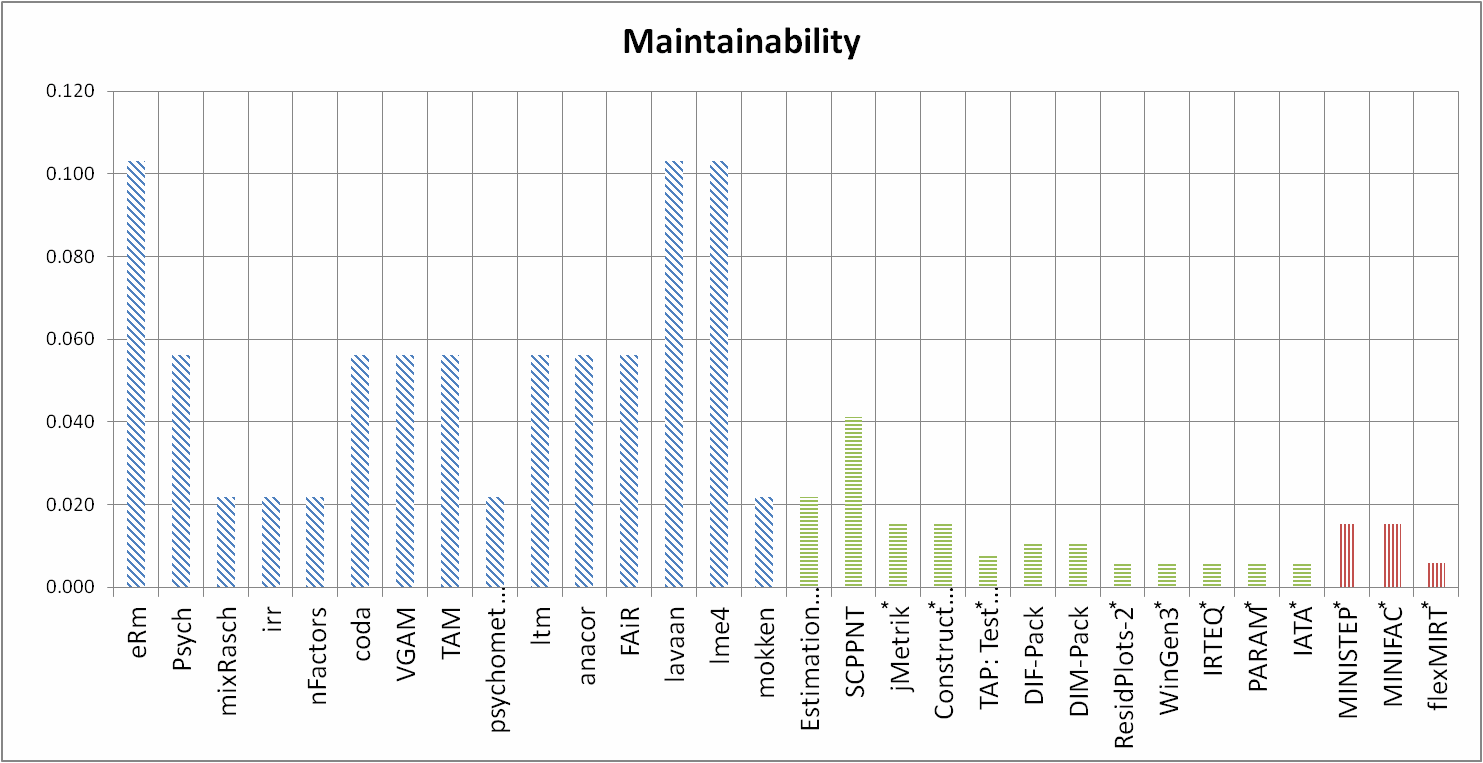}
\caption{AHP result for maintainability (close source indicated by *)}
\label{fig:maintainability}  
\end{figure}

\begin{itemize}

\item \proglang{R} packages: All of them provide a history of multiple versions
  and give information about how the packages were checked.  A few of them, like
  \pkg{lavaan} and \pkg{lme4}, also give information about how to
  contribute. $9$ provide change logs, $4$ indicate use of an issue tracking
  tool (Tracker and GitHub) and versioning systems (SVN and GitHub).  All of
  them consider maintainability in the code, with no obvious evidence of code
  clones.

\item Research group projects: Research group projects did not show much
  evidence that they pay attention to maintainability. Only $5$ provide the
  version history of their software; two give information about how to
  contribute and three provided change logs. None of them showed evidence of
  using issue tracking tools, or of using a versioning system in their project.

\item Commercial software: Because of the nature of commercial software, they
  did not show much \emph{externally visible} evidence of maintainability.  Two
  did provide history of multiple versions of the software and change logs, but
  other information is usually not provided by the vendors.  In this case, our
  measurements may not be an accurate reflection of the maintainability of these
  packages.

\end{itemize}

\subsubsection{Reusability}

We checked to see if the given software is itself used by another package, or if
there is evidence that reusability was considered in the design.  With reference
to Figure~\ref{fig:reusability}, our observations are as follows:

\begin{figure}[ht!]
\centering
\includegraphics[width=0.9\textwidth]{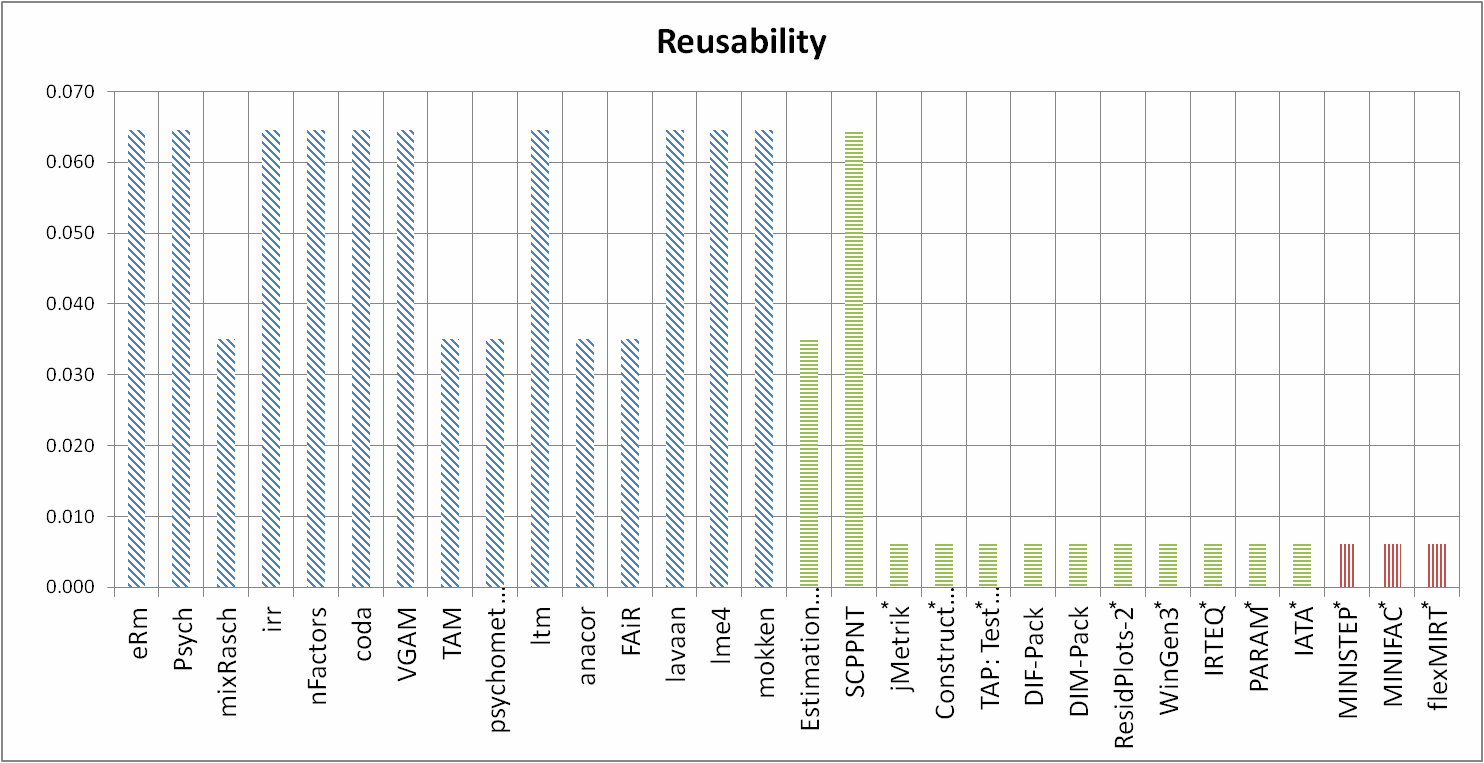}
\caption{AHP result for reusability (close source indicated by *)}
\label{fig:reusability}
\end{figure}

\begin{itemize}

\item \proglang{R} packages: There is clear evidence that 12 out of 15 software
  packages have been reused by other packages.  Also, the \proglang{Rd}
  reference manual required by CRAN can serve as an API document (of sorts),
  which helps others to reuse the package.

\item Research group projects: No evidence was found that their packages have
  been reused.  Two of them (\pkg{ETIRM} and \pkg{SCPPNT}) provide API
  documentation.

\item Commercial software: Due to their nature, no accurate result can be
  presented here.

\end{itemize}
 
\subsubsection{Portability}

We checked which platforms the software is advertised to work on, how people are
expected to handle portability, and whether there is evidence in the
documentation which shows that portability has been achieved.  Since the results
are so uniform by category, we omit the figure.

\begin{itemize}

\item \proglang{R} packages: These are consistently portable.  There are
  different versions of the package for different platform. Also, software
  checking results from CRAN showed that it does work well on different
  platforms.
\item Research group projects: All claim to work on Windows, with little
  evidence of working on other platforms.  An attempt for portability does not
  appear to have been made.
\item Commercial software: The results are similar to the research group
  projects.
\end{itemize}

\subsubsection{Understandability (Surface)}

We checked how easy it is for people to understand the code.  We focused on code
formatting style, existence of the use of coding standards and comments,
identifier and parameters style, use of constants, meaningful file names,
whether code is modularized and whether there is a design document.  The results
are summarized in Figure~\ref{fig:understandability}.  We can points out the
following:

\begin{figure}[ht!]
\centering
\includegraphics[width=0.9\textwidth]{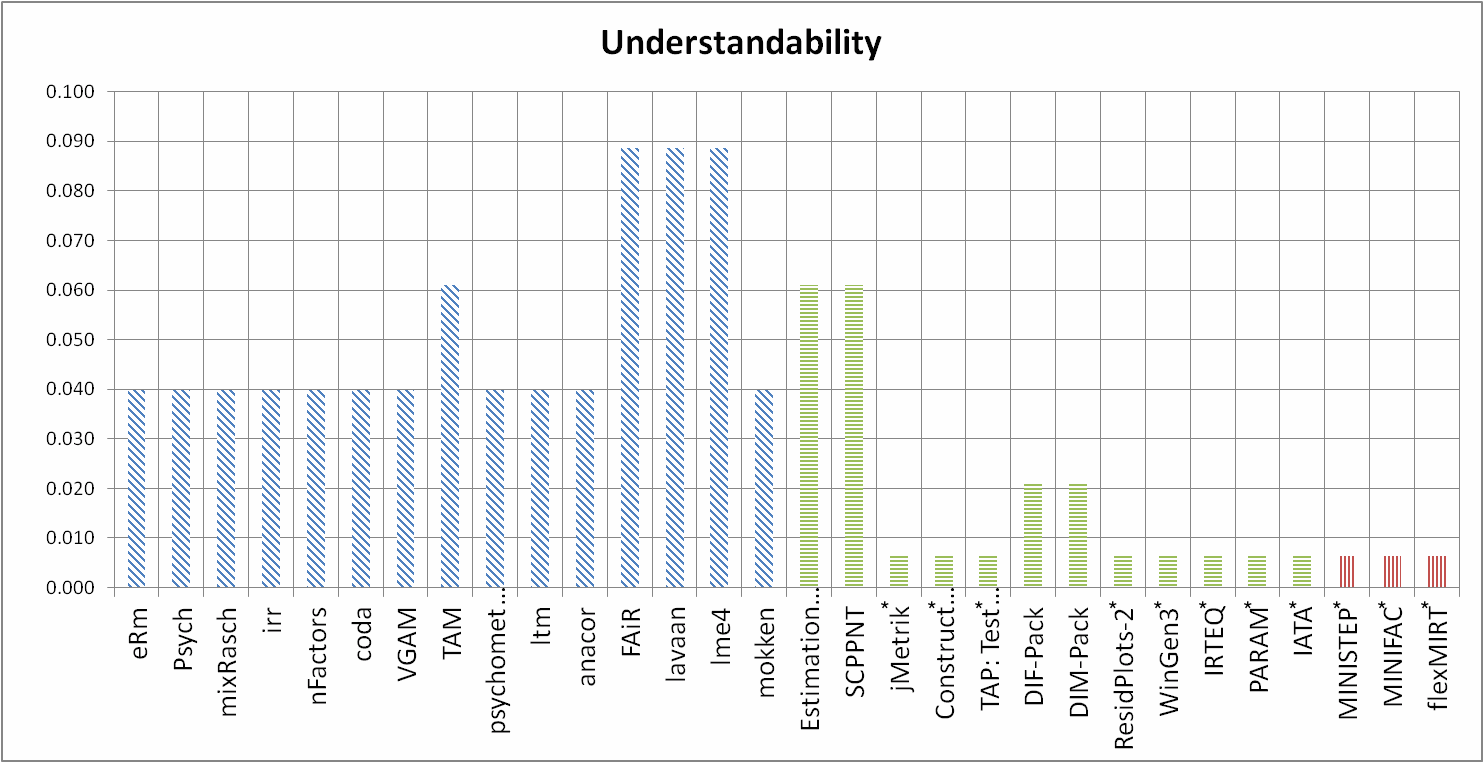}
\caption{AHP result for understandability (close source indicated by *)}
\label{fig:understandability}
\end{figure}

\begin{itemize}

\item \proglang{R} packages: These generally do quite well, with some specific
  stand-outs: (\pkg{FAiR}, \pkg{lavaan}, \pkg{lme4}).  One definite issue is
  that no coding standard is provided (for \proglang{R} in general) and there
  are few comments in the code.  If LP were used to maintain the code and its
  documentation together, understandability could be greatly improved.

\item Research group projects: 8 out of 12 projects are closed source.  For the
  closed source projects, no measure of understandability is given.  The common
  problem for the remaining open source projects is not specifying any coding
  standard.

\item Commercial software: Due to commercial software being closed source, no
  accurate result can be presented here.

\end{itemize}

\subsubsection{Interoperability}

We looked for evidence for any external systems the software communicates or
interoperates with, for a workflow that uses other software, and for external
interactions (API) being clearly defined.  The results for interoperability of
the SSP follow.  As the results are similar within each
category, we omit the figure.

\begin{itemize}

\item \proglang{R} packages: These do quite well, largely because of the
  documentation requirements from CRAN.  While there is no obvious evidence that
  these packages communicate with external system, it is very clear that
  existing workflows use other \proglang{R} packages.

\item No sign of interoperability is found in commercial software and research
  group projects.

\end{itemize}

\subsubsection{Visibility/Transparency}

\begin{figure}[ht]
\centering
\includegraphics[width=0.9\textwidth]{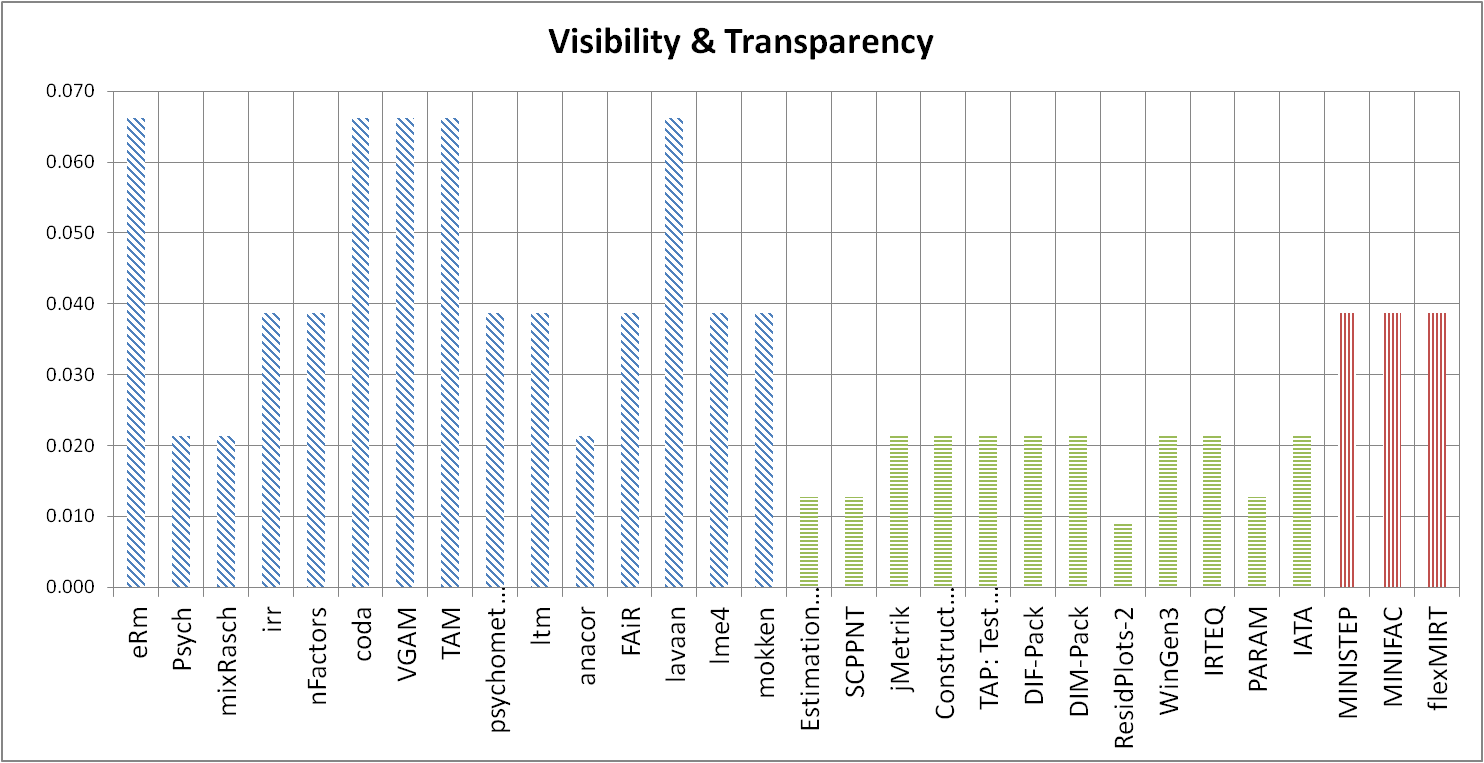}
\caption{AHP result for visibility/transparency}
\label{fig:visibility}
\end{figure}

We looked for a description of the development process.  We also give a
subjective score about the ease of external examination of the product, relative
to the average of the other products being considered.  This score is intend to
reflect how easy it is for other users to get useful information from the
software web site or documentation.  The AHP results are shown in
Figure~\ref{fig:visibility}.

\begin{itemize}

\item \proglang{R} packages: No information about the development process is
  given, but packages like \pkg{lavaan} provide web sites outside of CRAN with
  detailed information about the software, including examples and instructions.
  The benefit of the external web sites is that they provide more information to
  users; the downside is that maintaining information in two different places
  can be difficult.  If an external web site is used, developers need to make
  sure that all the web sites are up to date, and that links exist between them.

\item Research group projects and commercial software: None provide information
  about their development process for either of these categories, but
  information on the software is easily obtainable through web sites and
  documentation.

\end{itemize}

\subsubsection{Reproducibility}

We looked for a record of the environment used for development and testing, test
data for verification and automated tools used to capture experimental context.
Only \proglang{R} packages provide a record of the environment for testing (all
of them); the other products do not explicitly address reproducibility.
\proglang{R} packages also benefit from the use of \proglang{Sweave} for
vignettes (if present).

\section{Discussion}

It should be clear that different development groups have different priorities;
research projects typically aim to advance the state-of-the-art, while commercial
projects wish to be profitable.  This tends to mean that commercial projects have
more of an incentive to provide user-friendly software, with well written manuals,
and so on.  To a certain extent, the \proglang{R} community tends to fall in the
middle of these extremes.  On the other hand, it is not clear that such priorities
are always reflected in the actual development process.  The authors are well aware
of research software where very careful attention was paid to usability, and of
commercial software which is barely usable (but succeeds because it provides a 
service for which there is no effective competition).

In other words, while it may appear to be unfair to compare commercial and research
software, we disagree: assuming reasonable prices, users do not care so much 
about these details, but they do care about using software tools to accomplish their
tasks.  Being scientists, it is fair for us to measure these tools, even if it turns
out that the answer simply agrees with conventional wisdom.

Below, we provide some rankings, based on the measurements reported in the previous
section.

\begin{figure}[ht!]
\centering
\includegraphics[width=0.9\textwidth]{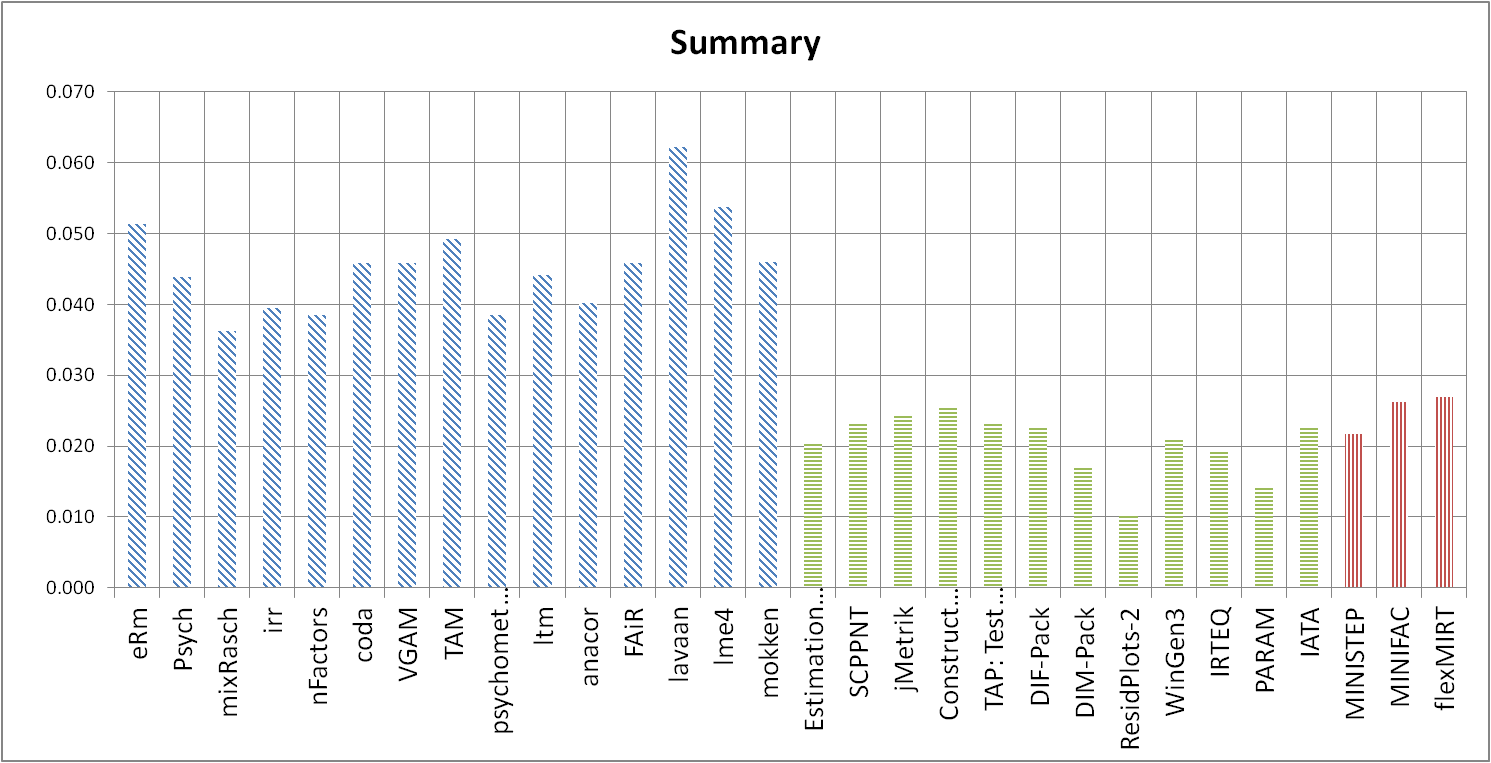}
\caption {Ranking with equal weight between all qualities}
\label{fig:nonweight}
\end{figure}

For our first analysis (see Figure~\ref{fig:nonweight}), we use the same weight
for each quality.  In this case, closed source packages ranked significantly
lower, since their open source counterparts provide more information for
qualities like maintainability, understandability, etc.  Thus, it is not
surprising that \proglang{R} packages fare dramatically better.  With respect to
research projects, the \proglang{R} packages may outperform them because the
\proglang{R} community explicitly targets user readiness.  As discussed in the
introductory section, some software developers implicitly target the ``lower''
goal of research readiness, since the burdens on design and documentation are
lower in this case.

\begin{figure}[ht!]
\centering
\includegraphics[width=0.9\textwidth]{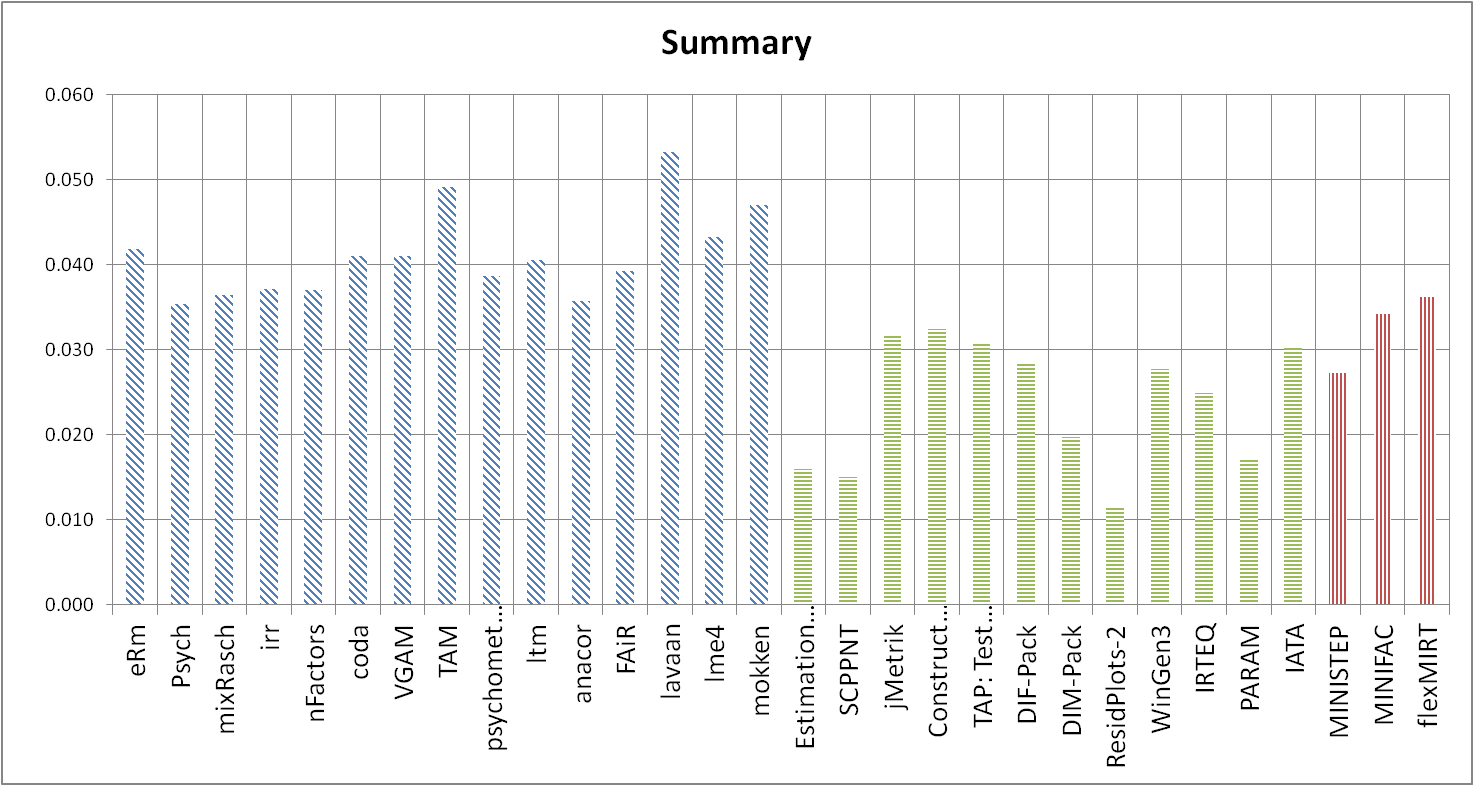}
\caption{Ranking with minimal weight for correctness \& verifiability,
  maintainability, reusability, understandability and reproducibility}
\label{fig:weight}
\end{figure}

To minimize the influence of open/closed source, we gave correctness \&
verifiability, maintainability, reusability, understandability and
reproducibility low weights (a weight of $1$ for these qualities, while all
others use $9$) -- See Figure~\ref{fig:weight}.  Interestingly, \proglang{R}
packages still fare best, but not to the same extent as in the previous
analysis.

One could argue that the results shown in Figure~\ref{fig:nonweight} are unfairly
biased against commercial software.  We disagree.  They are (unfairly?) biased
towards \emph{users} being able to ascertain that the product they have is of
high quality.  Most commercial software vendors make choices in how they package
their products which make this very difficult.  We believe that users should be
able to have confidence in the quality of their tools -- and to obtain such 
confidence through direct measurement.  This does mean that closed source is a
serious impediment; but this is not a barrier to commercialization, just to 
certain business models.

At the end of the day, every user must decide the relative weight that they
want to put on each quality, which is why our full data is available at
\url{https://github.com/adamlazz/DomainX}.

\section[Conclusion]{Conclusion and Recommendations} \label{conclusion}

For the surveyed software, \proglang{R} packages clearly performed far better
than the other categories for qualities related to development, such as
maintainability, reusability, understandability and visibility.  As we
expected, commercial software provided better usability but could not be
easily verified.  In terms of quality measurements, research
projects usually ranked lower and showed a larger variation in the quality
measures between products.

There is much to learn from the success of CRAN.  The overall high ranking of
\proglang{R} packages stems largely from their use of \proglang{Rd},
\proglang{Sweave}, \code{R CMD check} and the CRAN Repository Policy.  The
policy, together with the support tools, mean that even a single developer
project can be perceived externally as being sufficiently well developed to be
worth using.  A small research project usually does not have the resources to
set up an extensive development infrastructure and process, even though the end
results would benefit greatly from doing so.

As strong as the \proglang{R} community is, there is still room for improvement.
In particular, the documentation of an \proglang{R} extension seems to put too
much trust in the development team.  The documentation is solely aimed at
teaching the use of the code to a new user, not to convincing other developers
that the implementation is correct.  So while we applaud the use of the LP tool
\proglang{Sweave} for the user documentation, we are frankly puzzled that this
was not broadened to the code as well.  Another area for improvement would be an
increased usage of regression testing, supported by automated testing tools.
Regression testing can be used to ensure that updates have not broken something
that previously worked.

For each of the qualities, we can make further specific recommendations.  Some
of these sound incredibly obvious, but are nevertheless forced to make them as
we found sufficiently many instances of packages which did not do this.

\begin{itemize}

\item Installability: Systematically provide detailed installation instructions 
  and a standard suite of test cases for verification of the installation.

\item Correctness and Verifiability: Developers should consider following the
  example provided by CRAN, in terms of the organization of the reference
  manual, requirement specification, and information about the libraries used.
  With respect to user examples, commercial software, such as \pkg{flexMIRT},
  provides a wide variety of examples.  Although CRAN facilitates inclusion of
  examples, they are not generally required by repository policy -- they should
  consider increasing the number of required examples, possibly through
  vignettes, to match what is being done by commercial software.

\item Reliability: The user documentation, including installation instructions,
  need to be kept in sync with the software as it is updated.

\item Robustness: It was disappointing to see that several programs did not
  gracefully handle simple incorrect input.  Additional testing should be
  performed (and automated!), and issues uncovered, as shown in
  Appendix~\ref{app:SummaryMeasurements}, should be fixed.

\item Usability: CRAN should require a detailed getting started tutorial in
  addition to, or as part of, the user manual.  Commercial software should 
  put more effort in designing a platform-friendly GUI.

\item Maintainability: For open source projects, a versioning system and issue
  tracking tool are strongly suggested.  Information like a change log, or how
  to contribute, should also be presented.

\item Reusability: For programs to be reusable, a well-documented API should be
  provided.  If the generation of the documentation can be automated, there is a
  better chance that this will be done, and that it will be in sync with the
  code.

\item Understandability: Where not currently given, coding standards and design
  documents should be provided.  Developers should consider using LP for code
  development, so that the code can be written with the goal of human
  understanding, as opposed to computer understanding.  While the \proglang{R}
  community already uses such tools, for others, they can look to tools such as
  \proglang{cweb}, \proglang{noweb}, etc.

\item Visibility and Transparency: Projects that anticipate the involvement of
  future developers should provide details about the development process
  adopted.

\item Reproducibility: Developers should explicitly track the environment used
  for development and testing, to make the software results more reproducible.
  Through the use of the \proglang{R} environment and \proglang{Sweave}, CRAN
  provides good examples for how to do this.  However, the benefit of
  \proglang{Sweave} (or similar tools) would be improved if all CRAN developers
  were required to write vignettes for the package documentation.

\end{itemize}

\bibliographystyle{spbasic}

\bibliography{ref}

\newpage
\appendix

\section{Full grading template}\label{app:full-template}

The table below lists the full set of measures that are assessed for each
software product.  The measures are grouped under headings for each quality, and
one for summary information.  Following each measure, the type for a valid
result is given in brackets.  Many of the types are given as enumerated sets.
For instance, the response on many of the questions is one of ``yes,'' ``no,''
or ``unclear.''  The type ``number'' means natural number, a positive integer.  The
types for date and url are not explicitly defined, but they are what one would
expect from their names.  In some cases the response for a given question is not
necessarily limited to one answer, such as the question on what platforms are
supported by the software product.  Case like this are indicated by ``set of''
preceding the type of an individual answer.  The type in these cases are then
the power set of the individual response type.  In some cases a superscript $^*$
is used to indicate that a response of this type should be accompanied by
explanatory text.  For instance, if problems were caused by uninstall, the
reviewer should note what problems were caused.  An (I) precedes the question
description when its measurement requires a successful installation.

\begin{longtable}{p{0.95\textwidth}}
  \caption{Grading Template}   \label{table:TemplateFull}\\
  \toprule
  \textbf{Summary Information}\\
  \midrule
  Software name? (string)\\
  URL? (url)\\
  Educational institution (string)\\
  Software purpose (string)\\
  Number of developers (number)\\
  How is the project funded (string)\\
  Number of downloads for current version (number)\\
  Release date (date)\\
  Last updated (date)\\
  Status (\{alive, dead, unclear\})\\
  License (\{GNU GPL, BSD, MIT, terms of use, trial, none, unclear\})\\
  Platforms (set of \{Windows, Linux, OS X, Android, Other OS\})\\
  Category (\{concept, public, private\})\\
  Development model (\{open source, freeware, commercial\})\\
  Publications using the software (set of url)\\
  Publications about the software (set of url) \\
  Is source code available? (\{yes, no\})\\
  Programming language(s) (set of \{FORTRAN, Matlab, C, \CC, Java, R, Ruby,
  Python, Cython, BASIC, Pascal, IDL, unclear\})\\

  \midrule
  \textbf{Installability} (Measured via installation on a virtual machine.)\\
  \midrule

  Are there installation instructions? (\{yes, no\})\\
  Are the installation instructions linear? (\{yes, no, n/a\})\\
  Is there something in place to automate the installation? (\{yes$^*$, no\})\\
  Is there a specified way to validate the installation, such as a test suite? (\{yes$^*$, no\})\\
  How many steps were involved in the installation? (number)\\
  How many software packages need to be installed before or during installation?
  (number)\\
  (I) Run uninstall, if available. Were any obvious problems caused? (\{unavail, yes$^*$, no\})\\
  Overall impression? (\{1 .. 10\})\\

  \midrule
  \textbf{Correctness and Verifiability}\\
  \midrule

  Are external libraries used? (\{yes$^*$, no, unclear\})\\
  Does the community have confidence in this library? (\{yes, no, unclear\})\\
  Any reference to the requirements specifications of the program?
  (\{yes$^*$, no, unclear\})\\
  What tools or techniques are used to build confidence of correctness? (string)\\
  (I) If there is a getting started tutorial, is the output as expected? (\{yes, no$^*$, n/a\})\\
  Overall impression? (\{1 .. 10\})\\

  \midrule
  \textbf{Surface Reliability}\\
  \midrule

  Did the software ``break'' during installation? (\{yes$^*$, no\})\\
  (I) Did the software ``break'' during the initial tutorial testing? (\{yes$^*$, no, n/a\})\\
  Overall impression? (\{1 .. 10\})\\

  \midrule
  \textbf{Surface Robustness}\\
  \midrule

  (I) Does the software handle garbage input reasonably? (\{yes, no$^*$\})\\
  (I) For any plain text input files, if all new lines are replaced with new lines
  and carriage returns, will the software handle this gracefully? (\{yes,
  no$^*$, n/a\})\\
  Overall impression? (\{1 .. 10\})\\

  \midrule
  \textbf{Surface Performance}\\
  \midrule

  Is there evidence that performance was considered? (\{yes$^*$, no\})\\
  Overall impression? (\{1 .. 10\})\\

  \midrule
  \textbf{Surface Usability}\\
  \midrule

  Is there a getting started tutorial? (\{yes, no\})\\
  Is there a standard example that is explained? (\{yes, no\})\\
  Is there a user manual? (\{yes, no\})\\
  (I) Does the application have the usual ``look and feel'' for the platform it is
  on? (\{yes, no$^*$\})\\
  (I) Are there any features that show a lack of visibility? (\{yes, no$^*$\})\\
  Are expected user characteristics documented? (\{yes, no\})\\
  What is the user support model? (string)\\
  Overall impression? (\{1 .. 10\})\\

  \midrule
  \textbf{Maintainability}\\
  \midrule

  Is there a history of multiple versions of the software?  (\{yes, no, unclear\})\\
  Is there any information on how code is reviewed, or how to contribute?
  (\{yes$^*$, no\})\\
  Is there a changelog?  (\{yes, no\})\\
  What is the maintenance type? (set of \{corrective, adaptive, perfective, unclear\})\\
  What issue tracking tool is employed? (set of \{Trac, JIRA, Redmine, e-mail,
  discussion board, sourceforge, google code, git, none, unclear\})\\
  Are the majority of identified bugs fixed? (\{yes, no$^*$, unclear\})\\
  Which version control system is in use? (\{svn, cvs, git, github, unclear\})\\
  Is there evidence that maintainability was considered in the design?
  (\{yes$^*$, no\})\\
  Are there code clones? (\{yes$^*$, no, unclear\})\\
  Overall impression? (\{1 .. 10\})\\

  \midrule
  \textbf{Reusability}\\
  \midrule

  Are any portions of the software used by another package? (\{yes$^*$, no\})\\
  Is there evidence that reusability was considered in the design? (API
  documented, web service, command line tools, ...) (\{yes$^*$, no, unclear\})\\
  Overall impression? (\{1 .. 10\})\\

  \midrule
  \textbf{Portability}\\
  \midrule

  What platforms is the software advertised to work on?
  (set of \{Windows, Linux, OS X, Android, Other OS\})\\
  Are special steps taken in the source code to handle portability? (\{yes$^*$,
  no, n/a\}) \\
  Is portability explicitly identified as NOT being important? (\{yes, no\})\\
  Convincing evidence that portability has been achieved? (\{yes$^*$, no\})\\
  Overall impression? (\{1 .. 10\})\\

  \midrule
  \textbf{Surface Understandability} (Based on 10 random source files)\\
  \midrule

  Consistent indentation and formatting style? (\{yes, no, n/a\})\\
  Explicit identification of a coding standard? (\{yes$^*$, no, n/a\})\\
  Are the code identifiers consistent, distinctive, and
  meaningful? (\{yes, no$^*$, n/a\})\\
  Are constants (other than 0 and 1) hard coded into the program? (\{yes, no$^*$, n/a\})\\
  Comments are clear, indicate what is being done, not how? (\{yes, no$^*$, n/a\})\\
  Is the name/URL of any algorithms used mentioned?
  (\{yes, no$^*$, n/a\})\\
  Parameters are in the same order for all functions? (\{yes, no$^*$, n/a\})\\
  Is code modularized? (\{yes, no$^*$, n/a\})\\
  Descriptive names of source code files? (\{yes, no$^*$, n/a\})\\
  Is a design document provided? (\{yes$^*$, no, n/a\})\\
  Overall impression? (\{1 .. 10\})\\

  \midrule
  \textbf{Interoperability}\\
  \midrule

  Does the software interoperate with external systems? (\{yes$^*$, no\})\\
  Is there a workflow that uses other softwares? (\{yes$^*$, no\})\\
  If there are external interactions, is the API clearly defined? (\{yes$^*$, no, n/a\})\\
  Overall impression? (\{1 .. 10\})\\

  \midrule
  \textbf{Visibility/Transparency}\\
  \midrule

  Is the development process defined? If yes, what process is used. (\{yes$^*$, no, n/a\})\\
  Ease of external examination relative to other products
  considered?  (\{1 .. 10\})\\
  Overall impression? (\{1 .. 10\})\\

  \midrule
  \textbf{Reproducibility}\\
  \midrule

  Is there a record of the environment used for their development and testing?
  (\{yes$^*$, no\})\\
  Is test data available for verification?  (\{yes, no\})\\
  Are automated tools used to capture experimental context?  (\{yes$^*$, no\})\\
  Overall impression? (\{1 .. 10\})\\

  \bottomrule
\end{longtable}

\newpage

\section{Summary of Measurements} \label{app:SummaryMeasurements}

\subsection {Installability}
The columns of the table below should be read as follows:\\
II: Installation instructions available, II linear:
  Linear installation instructions, AI: Automated Installation,
  Inst Valid: Tests for installation validation, \# S/w lib: Number of
  software/libraries required for installation, uninst: Any uninstallation
  problem?

\renewcommand{\arraystretch}{0.6}
\begin{longtable}{lccccccc}
\label{TblInstallability}
\\
\toprule
Name & II & II linear  & AI & Inst Valid & \# of Steps & \# S/w lib & uninst
\\
\midrule 
\endhead
eRm & yes & yes & yes & no & 2 & 1 & no \\
Psych & yes & yes & yes & no & 4 & 4 & no \\
mixRasch & general & yes & yes & no & 2 & 1 & no \\
irr & general & yes & yes & no & 2 & 2 & no \\
nFactors & general & yes & yes & no & 2 & 4 & no \\
coda & general & yes & yes & no & 2 & 1 & no \\
VGAM & general & yes & yes & no & 2 & 4 & no \\
TAM & yes & yes & yes & no & 2 & 7 & no \\
psychometric & general & yes & yes & no & 2 & 3 & no \\
ltm & general & yes & yes & no & 2 & 3 & no \\
anacor & yes & yes & yes & no & 2 & 5 & no \\
FAiR & yes & yes & yes & no & 3 & 7 & no \\
lavaan & yes & yes & yes & yes & 2 & 13 & no \\
lme4 & yes & yes & yes & no & 2 & 16 & no \\
mokken & yes & yes & yes & no & 2 & 1 & no \\
ETIRM & yes & no & yes & no & 4 & 2 & n/a \\
SCPPNT & no & n/a & yes & no & 1 & 0 & n/a \\
jMetrik & yes & yes & yes & no & 1 & 0 & no \\
ConstructMap & yes & yes & yes & no & 1 & 1 & no \\
TAP & yes & yes & yes & no & 1 & 1 & no \\
DIF-Pack & yes & yes & yes & no & 5 & 1 & no \\
DIM-Pack & yes & yes & yes & no & 5 & 1 & no \\
ResidPlots-2 & no & n/a & yes & no & 1 & 1 & no \\
WinGen3 & yes & yes & yes & no & 2 & 2 & no \\
IRTEQ & no & n/a & yes & no & 2 & 2 & no \\
PARAM & no & n/a & yes & yes & 1 & 1 & no \\
IATA & no & n/a & yes & no & 2 & 2 & no \\
MINISTEP & yes & yes & yes & no & 1 & 1 & no \\
MINIFAC & yes & yes & yes & no & 1 & 1 & no \\
flexMIRT & yes & yes & yes & no & 1 & 1 & no \\
\bottomrule
\end{longtable}
\ifdefined\NONEWPAGE
   {}
\else
   \newpage
\fi

\subsection {Correctness and Verifiability}
The columns of the table below should be read as follows:\\
Library: Use of standard libraries,
SRS: Software Requirements Specification, Evidence?: Evidence to build
  confidence?, Example: Standard example explained.
\renewcommand{\arraystretch}{0.6}
\label{TblCorrectnessAndVerifiability}
\begin{longtable}{llccl}
\\
\toprule
Name & Library & SRS  & Evidence? & Example
\\
\midrule 
\endhead
eRm & yes (Cran)& yes & manual & yes \\
Psych & yes (Cran)& yes & manual & yes \\
mixRasch & yes (Cran)& no & manual & yes (no expected output) \\
irr & yes (Cran)& no & manual & yes (no expected output)\\
nFactors & yes (Cran)& no & manual & yes (no expected output)\\
coda & yes (Cran)& yes & manual & yes (no expected output)\\
VGAM & yes (Cran)& yes & manual & yes (no expected output)\\
TAM & yes (Cran)& no & manual & yes  \\
psychometric & yes (Cran)& no & manual & yes (no expected output)\\
ltm & yes (Cran)& yes & manual & small difference in answers\\
anacor & yes (Cran)& yes & manual & yes (no expected output)\\
FAiR & yes (Cran)& yes & manual & yes (no expected output)\\
lavaan & yes (Cran)& yes & manual & yes\\
lme4 & yes (Cran)& yes & manual & yes\\
mokken & yes (Cran)& yes & manual & yes\\
ETIRM & yes (SCPPNT) & no & no & yes (no expected output)\\
SCPPNT & no & no & no & yes  \\
jMetrik & yes (psychometric) & no & no & yes (not from latest
version) \\
ConstructMap & ? & no & test case & yes (output not clear) \\
TAP & ? & no & no & yes (no expected output) \\
DIF-Pack & ? & no & no & yes (no expected output)\\
DIM-Pack & ? & no & no & no\\
ResidPlots-2 & ? & no & no & Example data is not given\\
WinGen3 & ? & no & no & yes (no expected output)\\
IRTEQ & ? & no & no & yes\\
PARAM & ? & no & no & yes (no expected output)\\
IATA & ? & no & no & yes (no expected output)\\
MINISTEP & ? & no & no & yes\\
MINIFAC & ? & no & no & yes\\
flexMIRT & ? & no & no & yes\\
\bottomrule
\end {longtable}
\ifdefined\NONEWPAGE
   {}
\else
   \newpage
\fi

\subsection {Reliability, Robustness and Performance}
The columns of the table below should be read as follows:\\

install: Software breaks
    during installation, test: Software breaks during
    initial tutorial testing, Wrong I/P handling: Handling of wrong input by software, Format: Can the software
    gracefully handle a change of the format of text input files where the end of
    line follows a different convention?, Perf: Evidence that performance was
    considered?
\label {TblReliabilityEtc}
\renewcommand{\arraystretch}{0.6}
\begin {longtable} {l c c c c c}
\toprule
Name & install & test & Wrong I/P handling & Format & Perf \\
\midrule
\endhead
eRm & yes & no & yes & n/a & ? \\
Psych & yes & no & yes & n/a & ?\\
mixRasch & no & no & yes & n/a & ? \\
irr & no & no & ? & n/a & ? \\
nFactors & no & no & ? & n/a & ?  \\
coda & no & no & ? & n/a  & ? \\
VGAM & no & no & ? & n/a  & ? \\
TAM & no & no & yes & n/a & ?\\
psychometric & no & no & yes & n/a & ?\\
ltm & no & no & yes & n/a & ?\\
anacor & yes & no & ? & n/a & ?\\
FAiR & yes & yes & yes & n/a  & ?\\
lavaan & no & no & yes & n/a  & ? \\
lme4 & no & no & yes & n/a & ?\\
mokken & no & no & yes & n/a & ?\\
ETIRM & no & ? & ? & n/a & ? \\
SCPPNT & yes & no & yes & n/a & ?  \\
jMetrik & yes & no & yes & yes & ?  \\
ConstructMap & no & no & no & n/a & ? \\
TAP & no & no & yes & n/a & ?  \\
DIF-Pack & no & no & yes & n/a & ?\\
DIM-Pack & no & ? & yes & n/a & ?\\
ResidPlots-2 & no & ? & ? & n/a & ?\\
WinGen3 & no & no & yes & n/a & ?\\
IRTEQ & no & no & no & n/a & ?\\
PARAM & no & no & no & n/a & ?\\
IATA & no & no & yes & n/a & ?\\
MINISTEP & no & no & no & no & ?   \\
MINIFAC & no & no & yes & yes & ?  \\
flexMIRT & no & no & yes & n/a & ?\\
\bottomrule
\end {longtable}
\ifdefined\NONEWPAGE
   {}
\else
   \newpage
\fi

\subsection {Usability}
The columns of the table below should be read as follows:\\
Tut: Getting started tutorial, Ex: Standard example, UM:
User Manual, Look/feel: Usual look and feel of software, Vis: Lack of
visibility (Norman's Principle), User Char: User characteristics documented.
\label {TblUsability}
\renewcommand{\arraystretch}{0.6}
\begin {longtable} { l c c c c c c l}
\toprule
Name & Tut & Ex & UM & Look/feel & Vis & User Char & User
support \\
\midrule
\endhead
eRm & no & yes & yes & yes & yes & yes & Forum/Email \\
Psych & yes & yes & yes & yes & yes & no & Email \\
mixRasch & no & yes & yes & yes & yes & no & Email \\
irr & no & yes & yes & yes & yes & no & Email \\
nFactors & no & yes & yes & yes & yes & no & Email \\
coda & no & yes & yes & yes & yes & no & Email\\
VGAM & no & yes & yes & yes & yes & no & Email\\
TAM & yes & yes & yes & yes & yes & no & Email \\
psychometric & no & yes & yes & yes & yes & no & Email\\
ltm & no & yes & yes & yes & yes & no & Email/FAQ\\
anacor & no & yes & yes & yes & yes & no & Email/Forum\\
FAiR & no & yes & yes & yes & yes & no & Email\\
lavaan & yes & yes & yes & yes & yes & no & Email/Discussion group\\
lme4 & no & yes & yes & yes & yes & no & Email\\
mokken & yes & yes & yes & yes & yes & yes & Email\\
ETIRM & no & yes & no & no & yes & no & Email\\
SCPPNT & no & no & no & no & yes & no & Email\\
jMetrik & yes & yes & no & yes & no & yes & Email/FAQ/Tech support \\
ConstructMap & yes & yes & yes & yes & no & no & Email/FAQ\\
TAP & yes & yes & yes & yes & no & yes & Email \\
DIF-Pack & yes & yes & yes & no & no & no & Email/Discussion group \\
DIM-Pack & no & no & no & no & no & no & Email/Discussion group \\
ResidPlots-2 & no & yes & yes & no & no & no & Email \\
WinGen3 & yes & yes & yes & yes & no & no & Email/Survey \\
IRTEQ & yes & yes & yes & yes & no & no & Email \\
PARAM & no & no & yes & no & no & no & no \\
IATA & yes & yes & yes & no & no & yes & Email \\
MINISTEP & yes & yes & yes & no & no & no & Forum/Feedback/Email \\
MINIFAC & yes & yes & yes & no & no & no & Forum/Feedback/Email\\
flexMIRT & yes & yes & yes & yes & no & no & Email\\
\bottomrule
\end {longtable}
\ifdefined\NONEWPAGE
   {}
\else
   \newpage
\fi

\subsection {Maintainability}
The columns of the table below should be read as follows:\\
VH: Versions History available, RC: Information on reviewing and
  Contributing, log: Change log available, MT: Maintenance Type, Issue:
  Issue tracking tool, Bugs: Majority of bugs fixed, VS:
  Versioning System used, Evid: Any evidence that maintainability was
  considered in design, Clone: Are there code clones?, C: Corrective, P:
  Perfective, A: Adaptive.
\label {TblMaintainability}
\renewcommand{\arraystretch}{0.6}
\begin {longtable} {l c c c c c c c c c}
\toprule
Name & VH & RC & log & MT & Issue & Bugs & VS &
Evid & Clone\\
\midrule
\endhead
eRm & yes & yes & yes & C & Tracker & yes & svn & yes \\
Psych & yes & check & yes & C/P & ? & yes & ? & yes & no \\
mixRasch & yes & check & no & ? & ? & ? & ? & yes & no\\
irr & yes & check & no & ? & ? & ? & ? & yes & no \\
nFactors & yes & check & no & ? & ? & ? & ? & yes & no\\
coda & yes & check & yes & C/P & ? & yes & ? & yes & no   \\
VGAM & yes & check & yes & C/P & ? & yes & ? & yes & no\\
TAM & yes & check & yes & C/P & ? & yes & ? & yes & no \\
psychometric & yes & check & no & ? & ? & ? & ? & yes & no \\
ltm & yes & check & yes & C/P & ? & yes & ? & yes & no \\
anacor & yes & yes & no & ? & Tracker & yes & svn & yes & no\\
FAiR & yes & check & yes & C/P & ? & yes & ? & yes & no \\
lavaan & yes & yes & yes & C/P & git & yes & git & yes & no \\
lme4 & yes & yes & yes & C/P & git & yes & git & yes & no \\
mokken & yes & check & no & ? & no & ? & ? & yes & no \\
ETIRM & yes & no & no & ? & no & ? & ? & yes & no\\
SCPPNT & yes & no & yes & P & ? & yes & ? & yes & no \\
jMetrik & yes & no & yes & C/P & ? & ? & ? & ? & ?\\
ConstructMap & yes & no & yes & C/P & ? & ? & ? & ? & ?\\
TAP & yes & no & no & ? & ? & ? & ? & ? & ?\\
DIF-Pack & no & contri & no & ? & ? & ? & ? & no & no \\
DIM-Pack & no & contri & no & ? & ? & ? & ? & no & no\\
ResidPlots-2 & no & no & no & ? & ? & ? & ? & ? & ? \\
WinGen3 & no & no & no & ? & ? & ? & ? & ? & ?\\
IRTEQ & no & no & no & ? & ? & ? & ? & ? & ?\\
PARAM & no & no & no & ? & ? & ? & ? & ? & ?\\
IATA & no & no & no & ? & ? & ? & ? & ? & ?\\
MINISTEP & yes & no & yes & C/P & ? & ? & ? & ? & ? \\
MINIFAC & yes & no & yes & C/P & ? & ? & ? & ? & ?\\
flexMIRT & no & no & no & ? & ? & ? & ? & ? & ?\\
\bottomrule
\end{longtable}
\ifdefined\NONEWPAGE
   {}
\else
   \newpage
\fi

\subsection {Reusability}
\renewcommand{\arraystretch}{0.6}
\begin {longtable} {l c c}
\label {TblReusability} \\
\toprule
Name & S/w reused by any package & Any evidence of reusability \\
\midrule
\endhead
eRm & yes & yes (API)\\
Psych & yes & yes (API)\\
mixRasch & ? & yes (API)\\
irr & yes & yes (API)\\
nFactors & yes & yes (API)\\
coda & yes & yes (API)\\
VGAM & yes & yes (API)\\
TAM & yes & yes (API)\\
psychometric & yes & yes (API)\\
ltm & yes & yes (API)\\
anacor & no & yes (API)\\
FAiR & no & yes (API)\\
lavaan & yes & yes (API)\\
lme4 & yes & yes (API)\\
mokken & yes & yes (API)\\
ETIRM & no & yes\\
SCPPNT & yes & yes\\
jMetrik & no & no\\
ConstructMap & no & no\\
TAP & no & no\\
DIF-Pack & no & no\\
DIM-Pack & no & no\\
ResidPlots-2 & no & no\\
WinGen3 & no & no\\
IRTEQ & no & no\\
PARAM & no & no\\
IATA & no & no\\
MINISTEP & no & no\\
MINIFAC & no & no\\
flexMIRT & no & no\\
\bottomrule
\end {longtable}
\ifdefined\NONEWPAGE
   {}
\else
   \newpage
\fi

\subsection {Portability}
The columns of the table below should be read as follows:\\
Platforms: Platforms specified for the software to work on, Port in
  code: How portability is handled (if source code given), Port not imp:
  Portability explicitly identified as not important, Evid in doc: Convincing
  evidence present in documentation for portability?.
\label {TblPortability}
\renewcommand{\arraystretch}{0.6}
\begin {longtable} {l c c c l}
\toprule
Name & Platforms & Port in code  & Port not imp & Evid in doc \\
\midrule
\endhead
eRm & Win/Mac/Linux & Branch in repository & no & yes\\
Psych & Win/Mac/Linux & Branch in repository & no & yes\\
mixRasch & Win/Mac/Linux & Branch in repository & no & yes\\
irr & Win/Mac/Linux & Branch in repository & no & yes\\
nFactors & Win/Mac/Linux & Branch in repository & no & yes\\
coda & Win/Mac/Linux & Branch in repository & no & yes\\
VGAM & Win/Mac/Linux & Branch in repository & no & yes\\
TAM & Win/Mac/Linux & Branch in repository & no & yes\\
psychometric & Win/Mac/Linux & Branch in repository & no & yes\\
ltm & Win/Mac/Linux & Branch in repository & no & yes\\
anacor & Win/Mac/Linux & Branch in repository & no & yes\\
FAiR & Win/Mac/Linux & Branch in repository & no & yes\\
lavaan & Win/Mac/Linux & Branch in repository & no & yes\\
lme4 & Win/Mac/Linux & Branch in repository & no & yes\\
mokken & Win/Mac/Linux & Branch in repository & no & yes\\
ETIRM & Win/Mac/Linux & no & no & no\\
SCPPNT & Win/Mac/Linux & no & no & no\\
jMetrik & Win/Mac/Linux & n/a & no & ?\\
ConstructMap & Win/Mac/Linux & n/a & no & ?\\
TAP & Win & n/a & no & no\\
DIF-Pack & Win & ? & no & no\\
DIM-Pack & Win & ? & no & no\\
ResidPlots-2 & Win & n/a & no & no\\
WinGen3 & Win & n/a & no & no\\
IRTEQ & Win & n/a & no & no\\
PARAM & Win & n/a & no & no\\
IATA & Win & n/a & no & no\\
MINISTEP & Win & n/a & no & yes\\
MINIFAC & Win & n/a & no & yes\\
flexMIRT & Win & n/a & no & no\\
\bottomrule
\end {longtable}
\ifdefined\NONEWPAGE
   {}
\else
   \newpage
\fi

\subsection {Understandability (of code)}
The columns of the table below should be read as follows:\\
Frmt: Consistent identation and formatting, Std: Explicit
  coding standard, Id: Distinctive, meaningful Identifier names, Const:
  Constants (other than 0 or 1) hard coded, PC: Proper comments, Algo:
  Reference to algorithm used, Mod: Code is modularised, Para: Parameters
  are in same order, SC: Descriptive names of Source Code files, DD: Design
  Document present.
\label {TblUnderstandability}
\renewcommand{\arraystretch}{0.6}
\begin {longtable} {l l c c c c c c c c c}
\toprule
Name & Frmt & Std & Id & Const & PC & Algo & Mod & Para & SC & DD \\
\midrule
\endhead
eRm & yes & no & yes & no & no & no & yes & yes & yes & yes \\
Psych & yes & no & yes & no & no & no & yes & yes & yes & yes \\
mixRasch & yes & no & yes & no & no & no & yes & yes & yes & yes \\
irr & yes & no & yes & no & no & no & yes & yes & yes & yes \\
nFactors & yes & no & yes & no & no & no & yes & yes & yes & yes \\
coda & yes & no & yes & no & no & no & yes & yes & yes & yes \\
VGAM & yes & no & yes & no & no & no & yes & yes & yes & yes \\
TAM & yes & no & yes & no & yes & no & yes & yes & yes & yes \\
psychometric & yes & no & yes & no & no & no & yes & yes & yes & yes\\
ltm & yes & no & yes & no & no & no & yes & yes & yes & yes \\
anacor & no & no & yes & no & yes & yes & yes & yes & yes & yes \\
FAiR & yes & no & yes & no & yes & yes & yes & yes & yes & yes \\
lavaan & yes & no & yes & no & yes & yes & yes & yes & yes & yes \\
lme4 & yes & no & yes & no & yes & yes & yes & yes & yes & yes \\
mokken & yes & no & yes & no & yes & no & yes & yes & yes & yes \\
ETIRM & yes & no & yes & no & yes & no & yes & yes & yes & yes \\
SCPPNT & yes & no & yes & no & yes & no & yes & yes & yes & yes \\
jMetrik & n/a & n/a & n/a & n/a & n/a & n/a & n/a & n/a & n/a & no \\
ConstructMap & n/a & n/a & n/a & n/a & n/a & n/a & n/a & n/a & n/a & no \\
TAP & n/a & n/a & n/a & n/a & n/a & n/a & n/a & n/a & n/a & no \\
DIF-Pack & yes & no & yes & yes & yes & no & yes & no & yes & no \\
DIM-Pack & yes & no & yes & yes & yes & no & yes & no & yes & no \\
ResidPlots-2 & n/a & n/a & n/a & n/a & n/a & n/a & n/a & n/a & n/a & no \\
WinGen3 & n/a & n/a & n/a & n/a & n/a & n/a & n/a & n/a & n/a & no \\
IRTEQ & n/a & n/a & n/a & n/a & n/a & n/a & n/a & n/a & n/a & no \\
PARAM & n/a & n/a & n/a & n/a & n/a & n/a & n/a & n/a & n/a & no \\
IATA & n/a & n/a & n/a & n/a & n/a & n/a & n/a & n/a & n/a & no\\
MINISTEP & n/a & n/a & n/a & n/a & n/a & n/a & n/a & n/a & n/a & no \\
MINIFAC & n/a & n/a & n/a & n/a & n/a & n/a & n/a & n/a & n/a & no \\
flexMIRT & n/a & n/a & n/a & n/a & n/a & n/a & n/a & n/a & n/a & no \\
\bottomrule
\end{longtable}
\ifdefined\NONEWPAGE
   {}
\else
   \newpage
\fi

\subsection {Interoperability}
The columns of the table below should be read as follows:\\
External package: Software communicates with external package,
  Workflow uses other s/w: Workflow uses other software, API: External
  interactions (API) defined?
\label {TblInteroperability}
\renewcommand{\arraystretch}{0.6}
\begin {longtable} {l c c c}
\toprule
Name & External package & Workflow uses other s/w & API \\
\midrule
\endhead
eRm & no & yes & yes  \\
Psych & no & yes & yes\\
mixRasch & no & yes & yes\\
irr & no & yes & yes\\
nFactors & no & yes & yes\\
coda & no & yes & yes\\
VGAM & no & yes & yes\\
TAM & no & yes & yes\\
psychometric & no & yes & yes\\
ltm & no & yes & yes\\
anacor & no & yes & yes\\
FAiR & no & yes & yes\\
lavaan & no & yes & yes\\
lme4 & no & yes & yes\\
mokken & no & yes & yes\\
ETIRM & no & no & no  \\
SCPPNT & no & no & no\\
jMetrik & no & no & no\\
ConstructMap & no & no & no\\
TAP & no & no & no\\
DIF-Pack & no & no & no\\
DIM-Pack & no & no & no\\
ResidPlots-2 & no & no & no\\
WinGen3 & no & no & no  \\
IRTEQ & no & no & no\\
PARAM & no & no & no\\
IATA & no & no & no\\
MINISTEP & no & no & no\\
MINIFAC & no & no & no\\
flexMIRT & no & no & no\\
\bottomrule
\end{longtable}
\ifdefined\NONEWPAGE
   {}
\else
   \newpage
\fi

\subsection{Visibility/Transparency}
The columns of the table below should be read as follows:\\
  Dev process defined?: Development
    process defined, Ease of ext exam:
    Ease of examination relative to other software (out of 10).
\label {TblVisibility}
\renewcommand{\arraystretch}{0.6}
\begin{longtable} {l c c}
\toprule
Name & Dev process defined? & Ease of ext exam \\
\midrule
\endhead
eRm & no & 9  \\
Psych & no & 8   \\
mixRasch & no & 9  \\
irr & no & 9  \\
nFactors & no & 9\\
coda & no & 10   \\
VGAM & no & 10  \\
TAM & no & 10   \\
psychometric & no & 9  \\
ltm & no & 9   \\
anacor & no & 8  \\
FAiR & no & 7  \\
lavaan & no & 10  \\
lme4 & no & 9   \\
mokken & no & 9  \\
ETIRM & no & 6  \\
SCPPNT & no & 6   \\
jMetrik & no & 7   \\
ConstructMap & no & 7  \\
TAP & no & 6  \\
DIF-Pack & no & 6  \\
DIM-Pack & no & 6  \\
ResidPlots-2 & no & 5 \\
WinGen3 & no & 7  \\
IRTEQ & no & 7  \\
PARAM & no & 6  \\
IATA & no & 7  \\
MINISTEP & no & 8   \\
MINIFAC & no & 8\\
flexMIRT & no & 8\\
\bottomrule
\end{longtable}
\ifdefined\NONEWPAGE
   {}
\else
   \newpage
\fi

\subsection {Reproducibility}
The columns of the table below should be read as follows:\\
Dev env record: Record of development environment, Test data for
  verification: Availability of test data for verification, Auto tools: Automated
  tools (like Madagascar) used to capture experimental data.
\label {TblReproducibility}
\renewcommand{\arraystretch}{0.6}
\begin{longtable} {l c c l}
\toprule
Name & Dev env record & Test data for verification & Auto tools \\
\midrule
\endhead
eRm & only for testing & no & no  \\
Psych & only for testing & no & no\\
mixRasch & only for testing & no & no\\
irr & only for testing & no & no\\
nFactors & only for testing & no & no\\
coda & only for testing & no & no\\
VGAM Snap & only for testing & no & no\\
TAM & only for testing & no & no\\
psychometric & only for testing & no & no\\
ltm & only for testing & no & no\\
anacor & only for testing & no & no\\
FAiR & only for testing & no & no\\
lavaan & only for testing & no & no\\
lme4 & only for testing & no & no\\
mokken & only for testing & no & no\\
ETIRM & no & no & no  \\
SCPPNT & no & no & no\\
jMetrik & no & no & no\\
ConstructMap & no & no & no\\
TAP & no & no & no\\
DIF-Pack & no & no & no\\
DIM-Pack & no & no & no\\
ResidPlots-2 & no & no & no\\
WinGen3 & no & no & no  \\
IRTEQ & no & no & no  \\
PARAM & no & no & no  \\
IATA & no & no & no\\
MINISTEP & no & no & no\\
MINIFAC & no & no & no\\
flexMIRT & no & no & no\\
\bottomrule
\end{longtable}

\end{document}